\documentclass[9pt,twoside]{opticajnl}
\journal{opticajournal} 

\setboolean{shortarticle}{false}

\usepackage{lineno}

\usepackage{dcolumn}
\usepackage{bm, amssymb, color}
\usepackage{hyperref}
\usepackage{natbib}
\usepackage{booktabs,tabulary}
\usepackage{multirow}
\usepackage{amsmath,soul}
\usepackage[export]{adjustbox}
\usepackage{float}
\usepackage[caption=false]{subfig}
\usepackage{array}
\usepackage{cases}
\usepackage{makecell}
\usepackage{physics}
\usepackage{yhmath}

\newcommand{\spD}[1]{\fn{\tilde{\chi}_{_V}}{#1}}

\newcommand{\tens}[1]{\boldsymbol{#1}}

\newcommand{\uvect}[1]{\hat{\vect{#1}}}

\newcommand{\R}{\mathbb{R}}

\newcommand{\fn}[2]{\mathinner{#1\mathopen{\left(#2\right)}}}
\newcommand{\vect}[1]{{\bf #1}}
\newcommand{\E}[1]{\left\langle#1\right\rangle}

\newcommand{\F}[2]{\fn{F^\mathrm{(#1D)}}{#2}}
\newcommand{\J}[2]{\fn{\mathcal{J}^{(#1)}}{#2}}
\newcommand{\FTJ}[2]{\fn{\tilde{\mathcal{J}}^{(#1)}}{#2}}
\newcommand{\BETA}[2]{\beta^\mathrm{(#1D)}_{#2}}

\newcommand{\CTM}[1]{C_{#1}^{TM}}
\newcommand{\CL}[1]{C_{#1}^{\perp}}

\newcommand{\supp}{\href{url}{Supplement 1}}

\newcommand{\ATM}[2]{\fn{A_{#1}^{TM}}{#2}}
\newcommand{\ATE}[2]{\fn{A_{#1}^{TE}}{#2}}
\newcommand{\ETM}[1]{\fn{\varepsilon_e^{TM}}{#1}}
\newcommand{\ETE}[1]{\fn{\varepsilon_e^{TE}}{#1}}

\usepackage{soul}
\setstcolor{red}

\setul{0}{0.3ex}

\usepackage[colorinlistoftodos,shadow,textwidth=18mm]{todonotes}

\usepackage{xr}

\begin{document}

\title{Theoretical Prediction of the Effective Dynamic Dielectric Constant of Disordered Hyperuniform Anisotropic Composites Beyond the Long-Wavelength
Regime}

\author[1,2,3]{Jaeuk Kim}
\author[1,2,3,4,*]{Salvatore Torquato}
\affil[1]{Department of Chemistry, Princeton University, Princeton, New Jersey 08544, USA}
\affil[2]{Princeton Materials Institute, Princeton University, Princeton, New Jersey 08544, USA}
\affil[3]{Department of Physics, Princeton University, Princeton, New Jersey 08544, USA}
\affil[4]{Program in Applied and Computational Mathematics, Princeton University, Princeton, New Jersey 08544, USA}
\affil[*]{torquato@princeton.edu}

\begin{abstract}
    Torquato and Kim [Phys. Rev. X {\bf 11}, 296 021002 (2021)] derived exact nonlocal strong-contrast expansions of the effective dynamic dielectric constant tensor $\fn{\tens{\varepsilon}_e}{\vect{k}_q,\omega}$ that treat general statistically anisotropic three-dimensional (3D) two-phase composite microstructures, which are valid well beyond the long-wavelength regime.
    Here, we demonstrate that truncating this general rapidly converging expansion at the two- and three-point levels is a powerful theoretical tool from which one can extract accurate approximations suited for various microstructural symmetries.
    Among other results, we show that such truncations yield closed-form formulas applicable to transverse polarization in layered media and transverse magnetic polarization in transversely isotropic media, respectively.
    We apply these formulas to estimate $\fn{\tens{\varepsilon}_e}{\vect{k}_q,\omega}$ for models of 3D disordered hyperuniform layered and transversely isotropic media: nonstealthy hyperuniform media and stealthy hyperuniform media.
    In particular, we show that stealthy hyperuniform layered and transversely isotropic media are perfectly transparent (trivially implying no Anderson localization, in principle) within finite wave number intervals through the third-order terms.
    For all models considered here, we validate that the second-order formulas, which depend on the spectral density, are already very accurate well beyond the long-wavelength regime by showing very good agreement with the finite-difference time-domain (FDTD) simulations.
    The high predictive power of the second-order formula is due to the fact that higher-order contributions are negligibly small, implying that it very accurately approximates multiple scattering through all orders.
    This implies that there can be no Anderson localization within the predicted perfect transparency interval in stealthy hyperuniform layered and transversely isotropic media in practice because the localization length (associated with only possibly negligibly small higher-order contributions) should be very large compared to any practically large sample size.
    Our predictive theory provides the foundation for the inverse design of novel effective wave characteristics of disordered and statistically anisotropic structures by engineering their spectral densities.  
\end{abstract}

\maketitle


\section{Introduction} \label{sec:intro}

It is increasingly being recognized that one can engineer a spectrum of correlated disordered patterns in composites/metamaterials and devices to achieve novel photonic and phononic characteristics with advantages over their periodic counterparts \cite{florescu_designer_2009,izrailev_anomalous_2012, man_isotropic_2013,ma_3d_2016,leseur_highdensity_2016,wu_effective_2017,xu_microstructure_2017,froufe-perez_band_2017,gkantzounis_freeform_2017, chen_designing_2018,gorsky_engineered_2019,   kim_multifunctional_2020, rohfritsch_impact_2020, yu_engineered_2021, romero-garcia_wave_2021,vynck_light_2021, kim_bragg_2022, sgrignuoli_subdiffusive_2022, 
granchi_nearfield_2022, tavakoli_65_2022, cheron_wave_2022, klatt_wave_2022, tang_hyperuniform_2023, shi_computational_2023}.
The concept of disordered hyperuniformity has changed our understanding of the nature of randomness and order \cite{torquato_local_2003,torquato_hyperuniform_2018} and has provided a means to quantify the degree of order/disorder across length scales \cite{torquato_local_2022}.
Hyperuniform systems have attracted great attention over the last decade because of their deep connections to a wide range of topics that arise in physics \cite{torquato_ensemble_2015,zhang_perfect_2016, hexner_enhanced_2017, oguz_hyperuniformity_2017,ma_random_2017,lopez_true_2018,yu_disordered_2018,wang_hyperuniformity_2018,torquato_hyperuniform_2018,lei_hydrodynamics_2019, gorsky_engineered_2019,klatt_cloaking_2020, nizam_dynamic_2021}, materials science \cite{ma_3d_2016,xu_microstructure_2017,torquato_multifunctional_2018,chen_designing_2018,kim_new_2019}, mathematics \cite{ghosh_generalized_2017,  brauchart_hyperuniform_2019,torquato_hidden_2019, lacroix_intermedidate_2019}, and biology \cite{jiao_avian_2014, mayer_how_2015} as well as for their emerging technological importance in the case of the disordered varieties \cite{florescu_designer_2009, man_isotropic_2013,ma_3d_2016,leseur_highdensity_2016,froufe-perez_band_2017, wu_effective_2017,klatt_characterization_2018,gorsky_engineered_2019,klatt_wave_2022, kim_bragg_2022,froufe-perez_bandgap_2023, ong_control_2023}.
A hyperuniform two-phase composite in $d$-dimensional Euclidean space $\R^d$ is characterized by an anomalous suppression of volume-fraction fluctuations in the infinite-wavelength limit, i.e., the spectral density $\spD{k}$ obeys the condition \cite{zachary_hyperuniformity_2009,torquato_hyperuniform_2018}:
\begin{align}
    \lim_{\abs{\vect{k}}\to 0} \spD{\vect{k}} = 0.
    \label{eq:HU-condition}
\end{align}
Hyperuniform two-phase media encompass all periodic systems, many quasiperiodic media, and exotic disordered ones and thus generalizes our notions of long-range order to include exotic disordered varieties; see Ref. \cite{torquato_hyperuniform_2018} and references therein. 
Disordered hyperuniform systems lie between liquids and crystals; they are like liquids in that they are statistically isotropic without any Bragg peaks, and yet behave like crystals in the manner in which they suppress the large-scale density fluctuations \cite{torquato_local_2003,zachary_hyperuniformity_2009,torquato_hyperuniform_2018}.
Equivalently, a hyperuniform two-phase medium is characterized by a local volume-fraction variance $\fn{\sigma_V^2}{R}$ associated with a spherical window of radius $R$ that goes to zero asymptotically more rapidly than the inverse of the window volume, i.e., faster than $R^{-d}$.
There are several classes of hyperuniform two-phase media, distinct from nonhyperuniform media in which the variance goes to zero like $R^{-d}$ or slower (see section \ref{sec:back}).
One important subclass of such hyperuniform media is the disordered {\it stealthy} varieties in which $\spD{\vect{k}}=0$ for $0<\abs{\vect{k}}<K$ \cite{uche_constraints_2004, batten_classical_2008,zhang_ground_2015,torquato_ensemble_2015}, meaning that they completely suppress single scattering of incident radiation for these wavevectors \cite{batten_classical_2008,torquato_hyperuniform_2018}.
These exotic disordered systems exhibit novel electromagnetic wave transport properties, including high transparency in the optically dense regime, maximized absorption, and complete photonic band-gap formation \cite{florescu_designer_2009, man_isotropic_2013, leseur_highdensity_2016, zhang_transport_2016,froufe-perez_band_2017, torquato_multifunctional_2018,chen_designing_2018,gorsky_engineered_2019,zhou_ultrabroadband_2020,  kim_multifunctional_2020, sheremet_absorption_2020, yu_engineered_2021, granchi_nearfield_2022, tavakoli_65_2022, cheron_wave_2022, klatt_wave_2022}.

A vital electromagnetic wave characteristic is the effective dynamic dielectric constant tensor $\fn{\tens{\varepsilon}_e}{\vect{k}_q, \omega}$ for the incident radiation of wave vector $\vect{k}_q$ and frequency $\omega$.
Calculation of $\fn{\tens{\varepsilon}_e}{\vect{k}_q, \omega}$ is crucial for a wide range of applications, including remote sensing of terrain or vegetation \cite{stogryn_electromagnetic_1974,tsang_theory_1977, tsang_scattering_1981,sihvola_electromagnetic_1999}, understanding wave propagation through turbulent atmospheres \cite{tatarskii_effects_1971}, and designing optical metamaterials \cite{silveirinha_design_2007}.
While there have been many numerical and theoretical treatments to predict $\fn{\tens{\varepsilon}_e}{\vect{k}_q, \omega}$ for both statistically anisotropic \cite{rytov_electromagnetic_1956,
sjoberg_exact_2006, maurel_effective_2008, chebykin_nonlocal_2011, popov_operator_2016, merzlikin_homogenization_2020, wen_nonlocal_2021, 
siqueira_method_1996, silveirinha_design_2007, odeh_optical_2021} and isotropic \cite{keller_stochastic_1964,sihvola_electromagnetic_1999,ruppin_evaluation_2000, rechtsman_effective_2008} media, the preponderance of previous approximations are either microstructure independent or applicable in the long-wavelength (or \emph{quasistatic}) regime, i.e., $\abs{\vect{k}_q}\xi \ll 1$, where $\xi$ is a characteristic inhomogeneity length scale.

Torquato and Kim \cite{torquato_nonlocal_2021} recently derived exact general nonlocal strong-contrast expansions that are rational functions of the effective dielectric constant tensor $\fn{\tens{\varepsilon}_e}{\vect{k}_q, \omega}$ and which depend of two-phase composites on the microstructure via functionals of the $n$-point correlation functions $\fn{S_n ^{(i)}}{\vect{x}_1,\ldots,\vect{x}_n}$ for all $n$ (see section \ref{sec:back}) and exactly treat multiple scattering to all orders beyond the long-wavelength regime up through the intermediate-wavelength regime (i.e., $0\leq \abs{\vect{k}_q}\xi \lesssim 1$).
Our strong-contrast formulas assume the phases are dissipationless, and hence, attenuation can only occur due to multiple-scattering effects.
While standard perturbation expansions of $\fn{\tens{\varepsilon}_e}{\vect{k}_q, \omega}$ \cite{keller_stochastic_1964, frisch_probabilistic_1968,  caze_diagrammatic_2015} do not converge rapidly for large contrast ratios, this linear fractional expansion rapidly converges well beyond the long-wavelength regime even when the phase contrast ratio is large, which is analytically demonstrated in Sec. S5 of \supp. 
Thus, truncation of this expansion at low order yields highly accurate estimates of $\fn{\tens{\varepsilon}_e}{\vect{k}_q, \omega}$ because higher-order contributions are negligibly small, implying that this resulting formula very accurately approximates multiple scattering through all orders \cite{torquato_nonlocal_2021,kim_effective_2023}.
Specifically, second-order truncations already provide accurate approximations beyond the long-wavelength regime for transverse electric (TE) polarization in transversely isotropic media \cite{torquato_nonlocal_2021} and transverse polarizations in 3D layered \cite{kim_effective_2023} and 3D fully isotropic media \cite{torquato_nonlocal_2021}.

In a more recent paper \cite{kim_effective_2023}, we focused on the application of the strong-contrast formalism to layered media, consisting of infinite parallel dielectric slabs of phases 1 and 2 whose thicknesses are derived from one-dimensional (1D) disordered media.
There, we showed, among other results, that the imaginary part of the effective dielectric constant at the two-point level is proportional to a sum involving both forward scattering and backscattering contributions, i.e., 
\begin{align}   \label{eq:im-layered-2pt}
    \Im[\fn{\varepsilon_e}{k_q}] 
    \propto 
    k_q \qty[\spD{0} + \spD{2k_q \sqrt{\E{\varepsilon}/\varepsilon_q}}],
\end{align}
where $\E{\varepsilon}\equiv \phi_1 \varepsilon_1 + \phi_2 \varepsilon_2$ is the arithmetic mean of the local dielectric constant, and $k_q$ is the wave number in the reference phase $q$ with dielectric constant $\varepsilon_q$; see definitions in section \ref{sec:theory}.
Thus, we see that a finite perfect transparency interval, i.e., 
\begin{align}   \label{eq:interval}
   0\leq k_q <K_T\equiv K/\qty(2\sqrt{\E{\varepsilon}/\varepsilon_q}),
\end{align}
can only be achieved only if the layered medium is both hyperuniform and stealthy in order for the forward scattering term [i.e., $\spD{0}$] and the backscattering term [i.e., $\spD{2k_q \sqrt{\E{\varepsilon}/\varepsilon_q}}$] to vanish.
For example, we note that a nonstealthy hyperuniform medium [i.e., $\spD{0}=0$ only] cannot have such a finite perfect transparency interval.

\begin{figure}
    \includegraphics[width=0.45\textwidth]{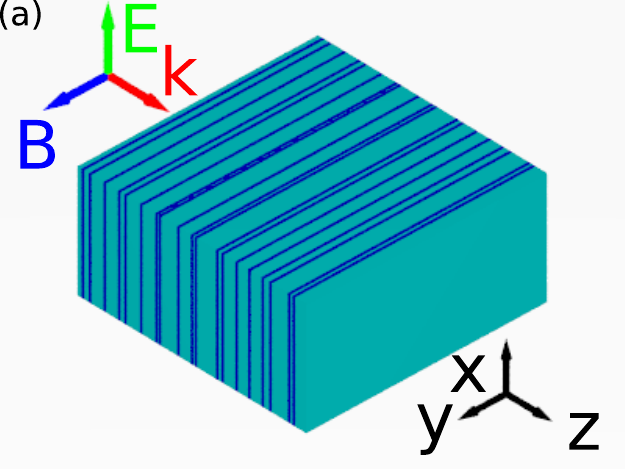}
    \includegraphics[width=0.45\textwidth]{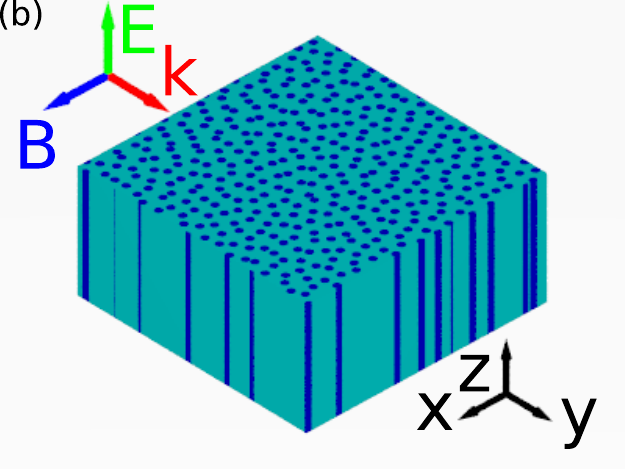}
    \caption{
        Schematic of 3D statistically anisotropic disordered two-phase media with two types of symmetries.
        We assume that both phases (shown in cyan and dark blue) of composites are dissipationless, and attenuation in these composites occurs only due to scattering.
        (a) Layered media consisting of infinite parallel slabs of phases 1 (cyan) and 2 (dark blue). 
        A plane electromagnetic wave of the transverse polarization is normally incident to the medium, and its wave vector is shown as a red arrow.
        (b) Transversely isotropic media with cylindrical symmetry derived from 2D isotropic two-phase media.
        A plane electromagnetic wave of transverse magnetic (TM) polarization is normally incident to the medium, and its wave vector is shown as a red arrow.
        \label{fig:schematics}
    }    
\end{figure}

This paper aims to demonstrate the versatility of the nonlocal strong-contrast expansion derived in Ref. \cite{torquato_nonlocal_2021} to provide accurate approximations suited for various structural symmetries by tuning the expansion parameter $\tens{L}_p^{(q)}$ (see Sec. S1 of \supp).
This task is carried out by choosing an appropriate exclusion volume shape around the singularity of the dyadic Green's function in the general expansion and then truncating the series at the $n$-point level, as briefly discussed in Ref. \cite{torquato_nonlocal_2021}.
We, for the first time, report the key predictive formulas of $\fn{\tens{\varepsilon}_e}{\vect{k}_q}$ extracted from the third-order truncations of the expansion for two types of statistically anisotropic two-phase composites: 3D layered media and 3D {\it transversely isotropic} media; see section \ref{sec:theory}.
For layered media, we focus on normally incident waves of transverse polarization, where $\vect{k}_q = k_q \uvect{z}$; see Fig. \ref{fig:schematics}(a).
Waves propagating in such anisotropic 3D media can be rigorously considered to be waves propagating in 1D ($d=1$) systems, and so we sometimes refer to them as 1D media.
The resulting third-order formula [\eqref{eq:eps-eff-strat_perp}] for layered media is a higher-order variant of the second-order formula derived in Ref. \cite{kim_effective_2023}.
Transversely isotropic media consist of infinite parallel dielectric cylinders of phase 2 embedded in phase 1 whose cross-section is derived from two-dimensional (2D) statistical isotropic media.
Here, we consider normally incident waves (i.e., $\vect{k}_q = k_q \uvect{y}$) and focus on transverse magnetic (TM) polarization; see Fig. \ref{fig:schematics}(b).
Such waves propagating in transversely isotropic 3D media can be rigorously considered to be waves propagating in 2D ($d=2$) systems and so we sometimes refer to them as 2D  media.
The resulting formula [\eqref{eq:eps-eff-2D_TM}] is for TM polarization, whereas the 2D formula in Ref. \cite{torquato_nonlocal_2021} is for TE polarization.
In both layered and transversely isotropic media, the wave number $k_q$ in the reference phase is the independent variable of the effective dielectric constant.

In the present work, we prove the perfect transparency interval predicted by the second-order formula [\eqref{eq:im-layered-2pt}] of stealthy hyperuniform layered media is also exact through third-order terms; see section \ref{sec:trans}.
We also prove that transversely isotropic stealthy hyperuniform media are perfectly transparent through the third-order terms for TM polarization within the interval given in \eqref{eq:interval}.
We observe, moreover, that the high predictive power of the second-order formula that predicts \eqref{eq:interval} trivially implies that there can be no Anderson localization within the predicted perfect transparency interval in 1D and 2D stealthy hyperuniform media in practice because the localization length, associated with only possibly negligibly small higher-order contribution, would be very large compared to any practically large sample size.

We also apply our strong-contrast formulas to estimate $\fn{\varepsilon_e}{k_q}$ for two models of disordered hyperuniform layered and transversely isotropic media: nonstealthy hyperuniform polydisperse packings in a matrix and stealthy hyperuniform packings in a matrix (see section \ref{sec:models}).
By contrast, Ref. \cite{kim_effective_2023} exclusively studied stealthy hyperuniform layered media.
For all models considered here, we numerically construct their realizations at $\phi_2=0.2$ and $0.25$ for layered media and transversely isotropic media, respectively.
For each model, we obtain the spectral density and then evaluate the nonlocal attenuation function, which is a crucial quantity in our second-order formulas.
Realizations of all models are later employed for the full-waveform simulations.

Subsequently, in section \ref{sec:results}, we corroborate the accuracy of the second-order formulas for both layered ($d=1$) and transversely isotropic ($d=2$) hyperuniform models for wave numbers ($k_1 \rho^{-1/d} \lesssim 1.5$) well beyond the long-wavelength regime by showing excellent agreement with the finite-difference time-domain (FDTD) simulations.
Furthermore, these approximations can provide qualitatively accurate predictions even beyond the intermediate-wavelength regime (i.e., $k_1 \rho^{-1/d} \gtrsim 1.5$) since they meet the Kramers-Kronig relations \cite{jackson_classical_1999} that relate the accurate low-frequency predictions to the high-frequency predictions.
Our second-order formulas predict that for hyperuniform models across dimensions, the real parts of the effective dielectric constants, closely related to the effective wave speed, similarly increase with $k_1$ for small wave numbers (i.e., $k_1\rho^{-1/d}<0.1$).
By contrast, these formulas predict that the two disordered models have qualitatively different attenuation behaviors, i.e., the imaginary parts of the effective dielectric constants.
However, they cannot achieve the finite perfect transparency intervals that stealthy hyperuniform models exhibit.
Since such accurate second-order formulas depend solely on the spectral density $\spD{\vect{k}}$, combining them with the methods to construct two-phase media with a targeted spectral density \cite{chen_designing_2018,shi_computational_2023,uche_constraints_2004,batten_classical_2008,zhang_ground_2015} would facilitate inverse approaches \cite{torquato_inverse_2009} to engineer and fabricate anisotropic dielectric media with tailored novel wave properties.

\section{Background}    \label{sec:back}
    \subsection{$n$-point correlation functions}
    \label{sec:n-point}

    A two-phase random medium is a domain of space $\mathcal V \subseteq \mathbb{R}^d$ that is partitioned into two disjoint regions that make up $\mathcal V$: a phase 1 region $\mathcal V_1$ of volume fraction $\phi_1$ and a phase 2 region $\mathcal V_2$ of volume fraction $\phi_2$ \cite{torquato_random_2002}. 
    The phase indicator function $\fn{\mathcal I^{(i)}}{\vect{x}}$ of phase $i$ for a given realization is defined as
    \begin{equation} \label{eq:indicator}
    \fn{\mathcal I^{(i)}}{\vect{x}}
    =
    \begin{cases}
        1, & \mathbf x \in \mathcal V_i \\
        0, & \text{otherwise}.
    \end{cases}
    \end{equation}
    The $n$-point correlation function $S_n^{(i)}$ for phase $i$ is defined as
    \cite{torquato_microstructure_1982,torquato_random_2002}
    $ \fn{S_n^{(i)}}{\mathbf{x}_1, \ldots, \mathbf{x}_n} = \E
    {\prod_{j=1}^n {\cal I}^{(i)}(\mathbf{x}_j)},
    $
    where $\E{\cdot}$ denotes an ensemble average over realizations. 
    The function $S_n^{(i)}(\mathbf{x}_1, \ldots, \mathbf{x}_n)$ has a probabilistic interpretation: It gives the probability of finding the ends of the vectors $\mathbf{x}_1, \ldots, \mathbf{x}_n$ all in phase $i$. 
    For statistically homogeneous media, $\fn{S_n^{(i)}}{\vect{x}_1, \ldots, \vect{x}_n}$ is translationally invariant
    and, in particular, the one-point function is position-independent, i.e., 
    $S_1^{(i)}(\mathbf{x_1}) = \phi_i$.

    The \textit{autocovariance} function $\chi_{_V}({\bf r})$, which is directly related to the two-point correlation function $\fn{S_2^{(i)}}{\vect{r}}$, is defined by  
    \begin{align}   \label{eq:spectral_density}
        \chi_{_V}({\bf r}) \equiv
        \E{\fn{\mathcal{J}^{(i)}}{\vect{x}_1} \fn{\mathcal{J}^{(i)}}{\vect{x}_1+\vect{r}} } 
        = S^{(i)}_2({\bf r}) - {\phi_i}^2,        
    \end{align}
    where $\fn{\mathcal{J}^{(i)}}{\vect{x}}\equiv \fn{\mathcal{I}^{(i)}}{\vect{x}}-\phi_i$ is the fluctuating part of \eqref{eq:indicator} for phase $i$.
    By definition, $\chi_{_V}({\bf r}=0)=\phi_1\phi_2$ and, assuming the medium possesses no long-range order, $\lim_{|{\bf r}| \rightarrow \infty} \chi_{_V}({\bf r})=0$.
    For statistically homogeneous and isotropic media, ${\chi}_{_V}({\bf r})$ depends only on the magnitude of its argument $r=\abs{\vect{r}}$, and hence is a radial function.
    The nonnegative spectral density $\spD{\vect{k}}$, which is proportional to scattering intensity \cite{debye_scattering_1957}, is the Fourier transform of $\fn{\chi_{_V}}{\vect{r}}$ at momentum-transfer wave vector $\vect{k}$.

    \subsection{Packings}   \label{sec:packing}

    We call a {\it packing} in $\R^d$ a collection of nonoverlapping particles \cite{torquato_perspective_2018}.
    For a packing in a periodic simulation box $\mathfrak{F}$, which consists of $N$ spheres of radii $a_1,\ldots,a_N$ and centers $\vect{r}_1,\ldots, \vect{r}_N$, its spectral density can be written as \cite{torquato_hyperuniformity_2016, torquato_perspective_2018}
    \begin{align}
        \spD{\vect{k}} 
        = \frac{1}{\abs{V_\mathfrak{F}}} {\abs{\FTJ{p}{\vect{k}}}^2}
        = \frac{1}{\abs{V_\mathfrak{F}}} {\abs{
            \sum_{j=1}^N \fn{\tilde{m}}{k;a_j} e^{-i \vect{k}\cdot \vect{r}_j} - \phi_2 \int_\mathfrak{F} \dd{\vect{r}} e^{-i\vect{k}\cdot\vect{r}} }^2
        },  \label{eq:spd-packing}
    \end{align}
    where $\abs{V_\mathfrak{F}}$ is the volume of $\mathfrak{F}$, $k\equiv \abs{\vect{k}}$, $\fn{\tilde{m}}{k; a} \equiv \qty(2\pi a/k)^{d/2} \fn{J_{d/2}}{ka}$, $\fn{J_\nu}{x}$ is the Bessel function of the first kind of order $\nu$, $\phi_2=\rho v_1(a)$ is the {\it packing
    fraction} (fraction of space covered by the spheres), and $v_1(a)= \pi^{d/2} a^d / \Gamma(1+d/2)$ is the $d$-dimensional volume of a sphere of radius $a$.
    In the case of a packing of identical spheres of radius $a$ at number density $\rho$ in $\R^d$, the spectral density $\spD{\vect{k}}$ is directly related to the structure factor $\fn{S}{\vect{k}}$ of the sphere centers \cite{torquato_random_2002, torquato_hyperuniformity_2016}:
    \begin{align}
        \spD{\vect{k}} = \rho \abs{\fn{\tilde{m}}{k; a}}^2 \fn{S}{\vect{k}}.    \label{eq:spd-spd-monodisperse}
    \end{align}

    \subsection{Classification of Hyperuniform and Nonhyperuniform Media} \label{sec:HU}

    The hyperuniformity concept generalizes the traditional notion of long-range order in many-particle systems to include not only all perfect crystals and perfect quasicrystals but also exotic amorphous states of matter, according to Refs. \cite{torquato_local_2003, torquato_hyperuniform_2018}.
    For disordered two-phase systems, the spectral density frequently exhibits a power-law scaling behavior in the small-wave number limit:
    \begin{align}   \label{eq:spd-power-law}
        \spD{\vect{k}} \sim \abs{\vect{k}}^\alpha   \qquad (\abs{\vect{k}} \to 0). 
    \end{align}
    The value of a positive exponent $\alpha$ relates to the three different scaling regimes (classes) associated with the large-$R$ behaviors of the volume-fraction variance \cite{zachary_hyperuniformity_2009,torquato_hyperuniform_2018}:
    \begin{align}  
    \fn{\sigma^2_{_V}}{R}
    \sim 
    \begin{cases}
        R^{-(d+1)},     & \qquad \alpha >1 \qquad \text{(Class I)},\\
        R^{-(d+1)} \ln R, & \qquad \alpha = 1 \qquad \text{(Class II)},\\
        R^{-(d+\alpha)}, & 0 < \alpha < 1\qquad  \text{(Class III)},
    \end{cases}
   \label{eq:HU-clases}
   \end{align}
   where the exponent $\alpha$ is a positive constant. 
   Classes I is the strongest form of hyperuniformity, which includes all perfect periodic media and some exotic disordered media.
   Stealthy hyperuniform media are also of class I and are defined as 
   \begin{align}    \label{eq:SHU-def}
        \spD{\vect{k}} = 0 \qquad \text{for } 0<\abs{k}<K.
   \end{align}

    By contrast, for any nonhyperuniform two-phase system, the variance has the following large-$R$ scaling behaviors \cite{torquato_structural_2021}:
    \begin{align}
        \fn{\sigma^2_{_V}}{R}
    \sim 
        \begin{cases}
            R^{-d}, &\quad~\quad~  \alpha = 0\quad \text{(typical nonhyperuniform)},\\
            R^{-d+\alpha},&-d<\alpha<0\quad\text{(antihyperuniform)},
        \end{cases} \label{eq:classes}
    \end{align}
    where $\alpha$ is defined in \eqref{eq:spd-power-law}.
    A typical nonhyperuniform system has a positive and finite $\spD{0}$, whereas an antihyperuniform one has an unbounded $\spD{0}$ that is diametrically opposite to hyperuniform systems.
    Antihyperuniform systems include systems at thermal critical points (e.g., liquid-vapor and magnetic critical points) \cite{stanley_introduction_1987, binney_theory_2002}, fractals \cite{mandelbrot_fractal_1982}, disordered nonfractals \cite{torquato_local_2022}, and certain substitution tilings \cite{oguz_hyperuniformity_2019}.


    \section{Theory}    \label{sec:theory}
    
    Here, we report the key formulas for the effective dynamic dielectric constant tensor of 3D layered media and 3D transversely isotropic media, which are extracted from the general strong-contrast expansion derived in Ref. \cite{torquato_nonlocal_2021}.
    Herein, we consider a plane electric wave of an angular frequency $\omega$ and a wavevector $\vect{k}_q$ in the reference phase $q$, which is taken as one phase (i.e., $q = 1$ or $2$) of the composite unless otherwise stated.
    We make three assumptions about both phases: (a) they have constant real-valued dielectric constants, (b) they are dielectrically isotropic, and (c) they are nonmagnetic.
    Thus, a dielectric composite that we consider here cannot absorb waves but attenuate them solely due to multiple scattering from fluctuations in the local dielectric constant.
    Since these assumptions give a simple relation between the wave number $k_q$ in the reference phase $q$ and angular frequency $\omega$ [i.e., $\fn{k_q}{\omega}\equiv\abs{\vect{k}_q(\omega)} = \sqrt{\varepsilon_q}\omega/c$], where $c$ is the speed of light in vacuum, we henceforth do not explicitly indicate the $\omega$ dependence, i.e., $\fn{\tens{\varepsilon}_e}{\vect{k}_q, \omega }=\fn{\tens{\varepsilon}_e}{\vect{k}_q}$.
    The linear fractional form of the strong-contrast expansion results in a rapidly converging series whose lower-order truncations lead to accurate approximation formulas for $\fn{\tens{\varepsilon}_e}{\vect{k}_q, \omega}$, even for large contrast ratios.
    This is to be contrasted with standard weak-contrast expansions that do not converge rapidly for large contrast ratios; see Sec. S5 of \supp~ for such a quantitative explanation.
    Indeed, its truncation at the two-point level yields a highly accurate estimation of $\fn{\tens{\varepsilon}_e}{\vect{k}_q, \omega}$ due to the fact that higher-order contributions are negligibly small, implying that this resulting formula very accurately approximates multiple scattering through all orders \cite{torquato_nonlocal_2021,kim_effective_2023}.
    Furthermore, similar to the strong-fluctuation theory \cite{tsang_scattering_1981,mackay_strongpropertyfluctuation_2000}, the expansion parameter of this series varies with the shape of the infinitesimal exclusion volume associated with the singularity of the dyadic Green's function.
    We exploit the fact that an appropriate choice of the exclusion volume can lead the series to be valid even for large values of phase contrast ratio  $\varepsilon_2/\varepsilon_1$ and various microstructural symmetries.
    We provide detailed derivations in Secs. S2 and S3 of \supp.


    \subsection{Multiple-Scattering Approximations for Layered Media}   \label{sec:layered}

    We consider our multiple-scattering approximations for the effective dynamic dielectric constant tensor of 3D layered media when waves are normally incident [see Fig. \ref{fig:schematics}(a)], i.e., $\vect{k}_q = k_q \uvect{z}$, where $\uvect{z}$ is a unit vector along the $z$-direction. 
    Thus, these formulas depend on the wave number $k_q$.
    They are extracted from exact strong-contrast series by choosing a disk-like exclusion volume normal to $\uvect{z}$ that involves the singularity of the dyadic Green's function and then truncating this series at the three-point level.
    We outline the derivation in Sec. S2 of \supp.

    Due to the symmetries of layered media, one can decompose the effective dielectric constant tensor into two orthogonal components $\fn{\varepsilon_e ^\perp}{k_q}$ and $\fn{\varepsilon_e ^z}{k_q}$ for the transverse and longitudinal polarizations, respectively, as follows:
    $\fn{\tens{\varepsilon}_e}{k_q} = \fn{\varepsilon_e ^\perp}{k_q} \qty(\tens{I}-\uvect{z}\uvect{z}) + \fn{\varepsilon_e^z}{k_q}\uvect{z}\uvect{z}$. 
    Thus, we derive two independent approximations by truncating the strong-contrast expansion and solving for $\fn{\tens{\varepsilon}_e}{k_q}$ from this linear fractional form [Eq. (S42) of \supp].
    Renormalization of the resulting formulas with the reference phase for the optimal convergence (Sec. S4 of \supp), equivalent to using the effective Green's function in Ref. \cite{torquato_nonlocal_2021}, yields \emph{scaled} strong-contrast approximations for disordered layered media at the three-point level:
    \begin{align}
        \fn{\varepsilon_e^\perp}{k_q} =& \varepsilon_q \qty[ 
            1 + \frac{{\phi_p}^2 \varepsilon_p/\varepsilon_q\BETA{1}{pq}}{\phi_p - (\varepsilon_p \BETA{1}{pq}) \fn{A_2^\perp}{k_q \sqrt{\E{\varepsilon}/\varepsilon_q}; \E{\varepsilon}} - (\varepsilon_p \BETA{1}{pq})^2 \fn{A_3^\perp}{k_q \sqrt{\E{\varepsilon}/\varepsilon_q}; \E{\varepsilon} } }
        ],
        \label{eq:eps-eff-strat_perp}
        \\
        \fn{\varepsilon_e^z}{k_q} =& \varepsilon_q (1 - \phi_p\BETA{1}{pq})^{-1},
        \label{eq:eps-eff-strat_z}
    \end{align}
    where $\BETA{1}{pq}\equiv  1-\varepsilon_q/\varepsilon_p$ is the one-dimensional counterpart of the \emph{dielectric polarizability}, and the second- and third-order terms are defined, respectively, as 
    \begin{align}
        \fn{A_2^\perp}{k_q;\varepsilon_q} 
        \equiv & \frac{1}{\varepsilon_q} \F{1}{k_q}
        = 
        \frac{1}{\varepsilon_q}\qty{\frac{{k_q}^2}{\pi} \mathrm{p.v.}\int_{0}^{\infty} \dd{q_z} 
        \frac{\spD{q_z}}{{q_z}^2 - {(2k_q)}^2}
        +
        \frac{i k_q}{4} [
        \spD{0} +  \spD{2k_q}
        ]}   \label{eq:F-strat-Fourier},    \\
        \fn{A_3^\perp}{k_q;\varepsilon_q} 
        \equiv & 
        \frac{-1}{\phi_p} \qty(\frac{{k_q}^2}{2\pi \varepsilon_q})^2
        \int_{-\infty}^\infty \dd{q_{1}} \frac{1}{(k_q+q_1)^2-{k_q}^2} \int_{-\infty}^\infty \dd{q_2} \frac{1}{(k_q+q_2)^2-{k_q}^2} \fn{\tilde{\Delta}_3^{(p)}}{q_1,q_2}
        \label{eq:A3-strat-Fourier},
    \end{align}
    where $\mathrm{p.v.}$ stands for the Cauchy principal value, and $\fn{\tilde{\Delta}_3^{(p)}}{q_1,q_2}$ is the Fourier transform of $\fn{\Delta_3^{(p)}}{z_{21},z_{31}}\equiv \fn{S_2^{(p)}}{z_{21}} \fn{S_2^{(p)}}{z_{32}} - \phi_p \fn{S_3^{(p)}}{z_{21}, z_{32}+z_{21}}$.
    Here, $\F{1}{k}$ is the {\it nonlocal attenuation function} for the layered media, and its two- and three-dimensional counterparts were derived in Ref. \cite{torquato_nonlocal_2021}.
    The second-order counterpart of \eqref{eq:eps-eff-strat_perp}, where $A_3^\perp = 0$, was derived in Ref. \cite{kim_effective_2023}, and its accuracy was numerically demonstrated there for stealthy hyperuniform layered media. 

    Note that $\fn{\varepsilon_e^\perp}{k_q}$ given in \eqref{eq:eps-eff-strat_perp} is complex-valued, implying that the media can be lossy due to forward scattering and backscattering from fluctuations in the local dielectric constant.
    By contrast, $\fn{\varepsilon_e^z}{k_q}$ is independent of $k_q$, reflecting the fact that a traveling longitudinal wave cannot exist under our assumptions.
    Hence, we focus on $\fn{\varepsilon_e^\perp}{k_q}$ in the rest of this work.   
    In the static limit, \eqref{eq:eps-eff-strat_perp} and \eqref{eq:eps-eff-strat_z} reduce to  the arithmetic and harmonic means of the local dielectric constants, respectively:
    \begin{align}
        \fn{\varepsilon_e^\perp}{0} =& \E{\varepsilon}\equiv \phi_p \varepsilon_p + \phi_q \varepsilon_q ,
        &\fn{\varepsilon_e^z}{0} =& (\phi_p/\varepsilon_p + \phi_q /\varepsilon_q)^{-1}.
        \label{eq:eps-eff-strat_perp-static}
    \end{align}
    Interestingly, these static results are exact for any microstructure  \cite{torquato_random_2002}.
    
    In the long-wavelength regime ($k_q /\rho \ll 1$), the imaginary part of \eqref{eq:eps-eff-strat_perp} is determined by the asymptotic behavior of $\fn{A_2^\perp}{k_q;\varepsilon_q}$.
    Thus, assuming that the spectral density has a power-law scaling [\eqref{eq:spd-power-law}], one immediately obtains 
    \begin{align}
        \Im[\fn{\varepsilon_e^\perp}{k_q}] 
        \sim (\varepsilon_p-\varepsilon_q)^2\Im[\fn{A_2^\perp}{k_q \sqrt{\E{\varepsilon}/\varepsilon_q}; \E{\varepsilon}}]
        \sim
        \begin{cases}
            k_q ,   & \alpha = 0 
            \quad ~~\text{(typical nonhyperuniform)} \\
            {k_q}^{1+\alpha}, & \alpha > 0
            \quad ~~\text{(hyperuniform)}
        \end{cases} .
        \label{eq:asymptotic-layered}
    \end{align}
    It is seen that hyperuniform media attenuate less than nonhyperuniform ones in the long-wavelength regime.
    Furthermore, while all typical nonhyperuniform systems exhibit a similar attenuation behavior [i.e., $\Im[\varepsilon_e]\propto {k_1}$ for small $k_1$], hyperuniform systems can exhibit a wide range of behaviors by tuning the exponent $\alpha$.
    The attenuation behavior of stealthy hyperuniform models is elaborated in section \ref{sec:trans}.

\subsection{Multiple-Scattering Approximations for Transversely Isotropic Media}    \label{sec:iso}

    We obtain our multiple-scattering approximations for the effective dynamic dielectric constant tensor of 3D transversely isotropic media for the situation in which waves are normally incident [see Fig. \ref{fig:schematics}(b)], i.e., $\vect{k}_q = k_q \uvect{y}$, where $\uvect{y}$ is a unit vector along the $y$-direction.
    These formulas depend on the wave number $k_q$ in the reference phase $q$.
    We extract them from the exact strong-contrast series by choosing a needle-like exclusion volume aligned along $\uvect{y}$ that involves the singularity of the dyadic Green's function and then truncating this series at the three-point level.
    We outline the derivation in Sec. S3 of \supp.

    Due to the symmetries of the problems, one can decompose the effective dielectric constant tensor into two orthogonal components $\fn{\varepsilon_e ^{TM}}{k_q}$ and $\fn{\varepsilon_e ^{TE}}{k_q}$ for TM and TE polarizations, respectively, as follows:
    $\fn{\tens{\varepsilon}_e}{k_q} = \fn{\varepsilon_e ^{TM}}{k_q} \uvect{z}\uvect{z} + \fn{\varepsilon_e^{TE}}{k_q}(\tens{I}-\uvect{z}\uvect{z})$.
    Thus, we extract two independent approximations of $\ETE{k_q}$ and $\ETM{k_q}$ by truncating the strong-contrast series at third-order terms; see Eq. (S78) and Eq. (S79) of \supp.
    Solving for the effective dielectric constants from these linear fractional forms and then renormalizing them with the optimal reference phase \cite{tsang_scattering_1981,mackay_strongpropertyfluctuation_2000, torquato_nonlocal_2021} (Sec. S4 of \supp), we finally obtain the \emph{scaled} strong-contrast approximations at the three-point level for disordered transversely isotropic media:
    \begin{align}
        \frac{\fn{\varepsilon_e^{TM}}{k_q}}{\varepsilon_q} 
        =& 
        1+ \frac{ 
            {\phi_p}^2 \qty[(\varepsilon_p+\varepsilon_q) \BETA{2}{pq}]
        } 
        {\phi_p - \ATM{2}{k_*^{TM}, \E{\varepsilon}} \qty[(\varepsilon_p+\varepsilon_q) \BETA{2}{pq}]
        - \ATM{3}{k_*^{TM}, \E{\varepsilon}} \qty[(\varepsilon_p+\varepsilon_q) \BETA{2}{pq}]^2 
        },    \label{eq:eps-eff-2D_TM}
        \\
        \frac{\ETE{k_q}}{\varepsilon_q}
        =& 
        1+ \frac{ 
            2  {\phi_p}^2 \BETA{2}{pq}
        }
        { 
            \phi_p(1-\phi_p \BETA{2}{pq}) 
            -  
            \ATE{2}{k_*^{TE}; \varepsilon_{BG}^{(2D)}} \qty[2\varepsilon_q \BETA{2}{pq}]
            -  
            \ATE{3}{k_*^{TE}; \varepsilon_{BG}^{(2D)}} \qty[2\varepsilon_q \BETA{2}{pq}]^2
        }
            \label{eq:2pt-trans-TE} , 
    \end{align}
    where $\BETA{2}{pq}\equiv (\varepsilon_p - \varepsilon_q)/(\varepsilon_p + \varepsilon_q)$ is the two-dimensional counterpart of the dielectric polarizability, and $k_*^{TE} \equiv k_q \sqrt{\varepsilon_{BG}^{(2D)}/\varepsilon_q}$ and $k_*^{TM} \equiv k_q \sqrt{\E{\varepsilon}/\varepsilon_q}$ are wave numbers in the optimal reference phase for TE and TM polarizations, respectively, and $\varepsilon_{BG}^{(2D)}$ is the Bruggeman approximation for 2D two-phase media \cite{bruggeman_berechnung_1935,torquato_random_2002}.
    We note that \eqref{eq:2pt-trans-TE} is identical to the 2D formula derived in Ref. \cite{torquato_nonlocal_2021}, except for the reference phase.
    The current choice for the reference dielectric constant $\varepsilon_{BG}^{(2D)}$ offers slightly better renormalization than 2D Maxwell-Garnett approximation employed in Ref. \cite{torquato_nonlocal_2021} because $\varepsilon_{BG}^{(2D)}$ makes the mean of the depolarization tensors exactly zero, i.e., $\phi_1 \tens{L}_1 ^{(*)} + \phi_2 \tens{L}_2 ^{(*)}=\tens{0}$ (see Eq. (S105) of \supp).
    Thus, we henceforth focus on the approximation for the TM polarization [\eqref{eq:eps-eff-2D_TM}].

    The second- and third-order coefficients of each polarization are defined as
    \begin{align}
        \fn{A_2^{TM}}{k_q; \varepsilon_q}
        =&
        2 \fn{A_2^{TE}}{k_q; \varepsilon_q}
        =
        - \frac{\pi}{2 \varepsilon_q} \F{2}{k_q} 
        = 
        \Biggl[
            \frac{{k_q}^2}{\varepsilon_q (2\pi)^2 }
            \int_{\R^2} \dd{\vect{q}}
            \frac{\spD{q}}{\abs{\vect{q} + \vect{k}_q}^2 - {k_q}^2} 
        \Biggr] 
        \label{eq:A2-trans}\\
        =&
        \frac{1}{\varepsilon_q}
        \qty{
            \frac{{k_q}^2}{\pi^2}
            \int_{0}^{\pi/2} \dd{\phi} 
            \qty[ \mathrm{p.v.} 
                \int_0^\infty \dd{q} \frac{2q\spD{q}}{q^2-(2k_q\cos\phi)^2}
            ]
        + i\frac{{k_q}^2}{\pi} \int_0^{\pi/2} \spD{2k_q \cos\phi}\dd{\phi}},
        \label{eq:F-trans-Fourier}
        \\
        \fn{A_3^{TM}}{k_q; \varepsilon_q}
        =& 
        \frac{-1}{\phi_p}\qty[\frac{{k_q}^2}{\varepsilon_q (2\pi)^2 }]^2
        \int_{\R^2} \dd{\vect{q}_1}
        \int_{\R^2} \dd{\vect{q}_2}
            \frac{1}{\abs{\vect{q}_1 + \vect{k}_q}^2 - {k_q}^2} 
            \frac{1}{\abs{\vect{q}_2 + \vect{k}_q}^2 - {k_q}^2} 
            \fn{\tilde{\Delta}_3^{(p)}}{\vect{q}_1,\vect{q}_2},
        \label{eq:A3-trans}
        \\
        \fn{A_3^{TE}}{k_q; \varepsilon_q}
        =& 
        \frac{-1}{\phi_p}
        \qty[\frac{1}{2\varepsilon_q (2\pi)^2}]^2
        \int_{\R^2} \dd{\vect{q}_1}
        \int_{\R^2} \dd{\vect{q}_2} 
        \frac{\Big[
            ({k_q}^2)^2 - \abs{\vect{k}_q+\vect{q}_1}^2 \abs{\vect{k}_q+\vect{q}_2}^2
            + 2 [(\vect{k}_q+\vect{q}_1)\cdot (\vect{k}_q+\vect{q}_2)]^2
        \Big]}
        {\qty(\abs{\vect{k}_q+\vect{q}_1}^2 - {k_q}^2) \qty(\abs{\vect{k}_q + \vect{q}_2}^2 - {k_q}^2)}
        \fn{\tilde{\Delta}_3^{(p)}}{\vect{q}_1,\vect{q}_2},
        \label{eq:A3-trans-TE}
    \end{align}
    where $\F{2}{k}$ is the nonlocal attenuation function for 2D statistically isotropic two-phase media \cite{torquato_nonlocal_2021}, and $\fn{\tilde{\Delta}_3^{(p)}}{\vect{q}_1,\vect{q}_2}$ is the Fourier transform of $\fn{\Delta_3^{(p)}}{\vect{x}_{21},\vect{x}_{31}}\equiv \fn{S_2^{(p)}}{\vect{x}_{21}} \fn{S_2^{(p)}}{\vect{x}_{32}} - \phi_p \fn{S_3^{(p)}}{\vect{x}_{21}, \vect{x}_{32}+\vect{x}_{21}}$.
    In the static limit ($k_q \to 0^+$), \eqref{eq:2pt-trans-TE} and \eqref{eq:eps-eff-2D_TM} reduce to the arithmetic means of the local dielectric constant and the third-order static strong-contrast approximation for $d=2$ \cite{torquato_random_2002}, respectively:
    \begin{align}
        \fn{\varepsilon_e^{TM}}{0} =& \E{\varepsilon} =\varepsilon_p\phi_p + \varepsilon_q \phi_q,    
        &\ETE{0} 
        =& 
        \varepsilon_q
        \frac{
            \phi_p (1+\phi_p\BETA{2}{pq}) - (1-\phi_p) \zeta_p [\BETA{2}{pq}]^2}
        {
            \phi_p (1-\phi_p\BETA{2}{pq}) - (1-\phi_p) \zeta_p [\BETA{2}{pq}]^2
        },    
        \label{eq:trans-TE-static} 
    \end{align}
    where $\zeta_p$ is the three-point parameter that lies in the closed interval $[0,1]$ \cite{torquato_random_2002}.

    In the long-wavelength regime ($k_q /\rho \ll 1$), the imaginary part of \eqref{eq:eps-eff-2D_TM} is determined by the asymptotic behavior of $\ATM{2}{k_*^{TM}; \E{\varepsilon}}$.
    Thus, assuming that the spectral density has a power-law scaling [\eqref{eq:spd-power-law}], one can immediately obtain 
    \begin{align}
        \Im[\ETM{k_q}] 
        \sim (\varepsilon_p-\varepsilon_q)^2\Im[\ATM{2}{k_*^{TM}; \E{\varepsilon}}]
        \sim
        \begin{cases}
            {k_q}^2 ,   & \alpha = 0 
            \quad ~~~\text{(typical nonhyperuniform)} \\
            {k_q}^{2+\alpha}, & \alpha > 0
            \quad ~~~\text{(hyperuniform)} 
        \end{cases} ,
        \label{eq:asymptotic-trans-TM}
    \end{align}
    which are identical to those for TE polarization \cite{torquato_nonlocal_2021}.
    Hyperuniform media attenuate less than nonhyperuniform ones in the long-wavelength regime.
    Furthermore, while all typical nonhyperuniform systems exhibit a similar attenuation behavior [i.e., $\Im[\varepsilon_e]\propto {k_1}^2$ for small $k_1$], hyperuniform systems can exhibit a wide range of behaviors by tuning the exponent $\alpha$. 
    The attenuation behavior of stealthy hyperuniform models is elaborated in section \ref{sec:trans}.

\begin{figure}[h]
    {\includegraphics[width=0.54\textwidth]{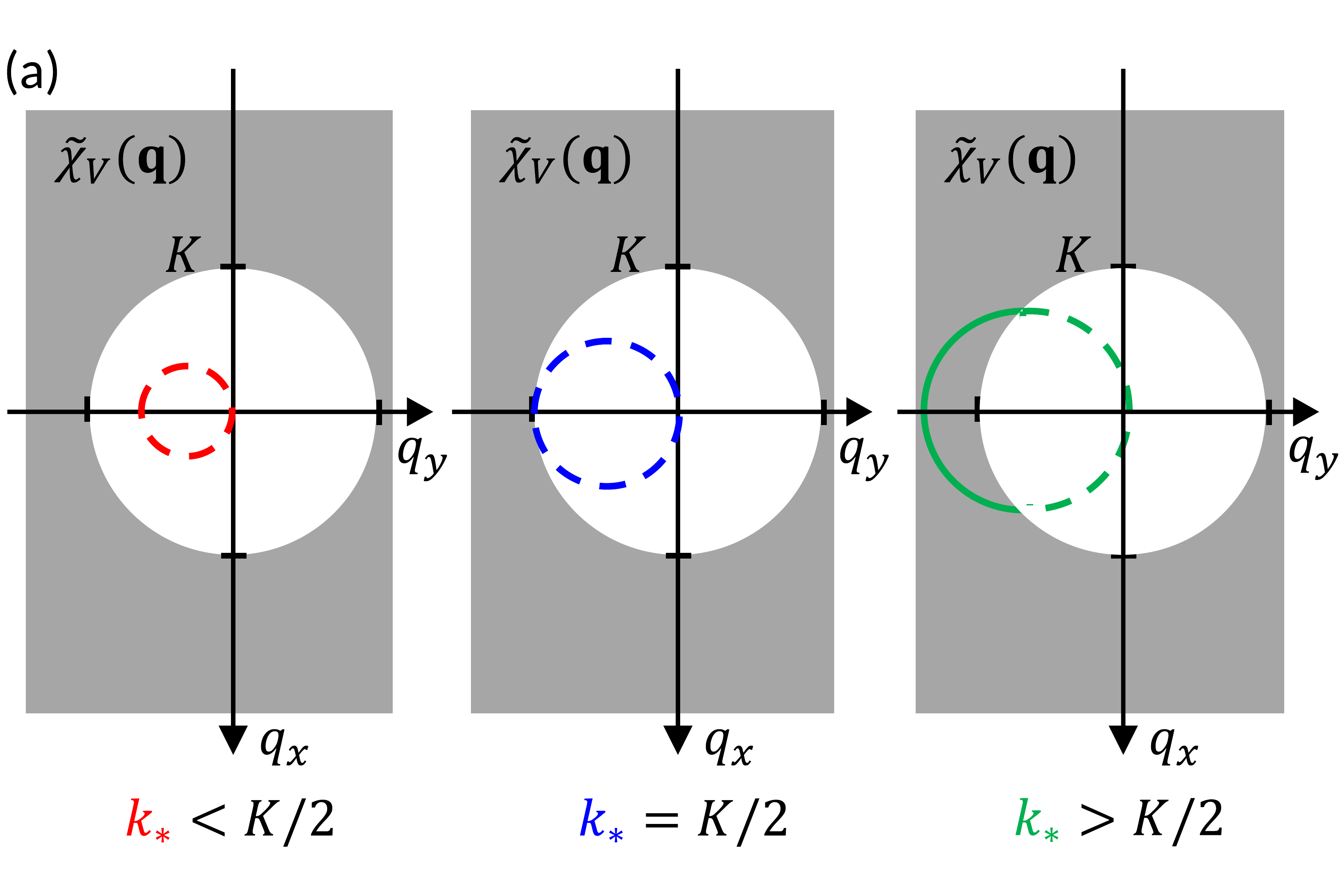}}
    \includegraphics[width=0.46\textwidth]{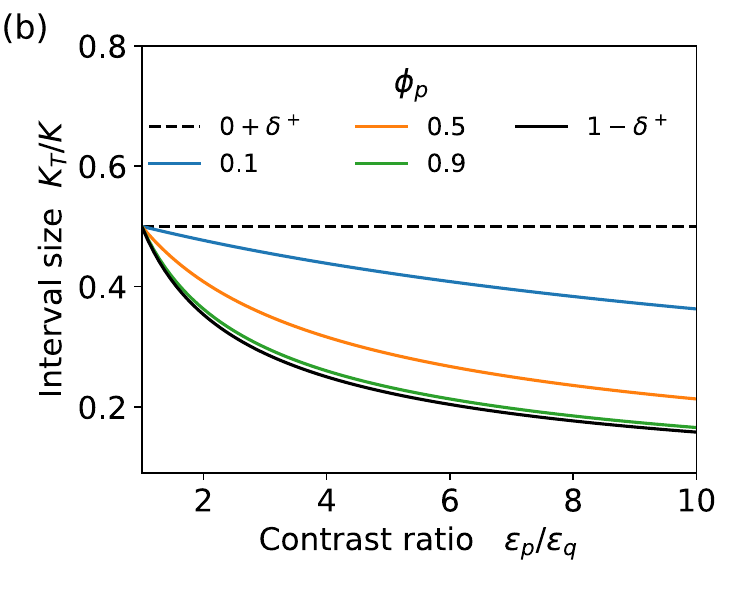}
    \caption{
    (a) Schematic of the on-shell scattering contributions to the nonlocal attenuation function for transversely isotropic stealthy hyperuniform media.
    The white and gray areas show the exclusion region of stealthy hyperuniform media and the rest of Fourier space, respectively.
    The colored circles indicate the on-shell wavevectors satisfying $\abs{k_*\uvect{y}+\vect{q}}=k_*$, in which elastic single-scattering event can occur.
    At the two-point level, since only on-shell scattering $\spD{\vect{q}}$ on such a circle contributes to attenuation, if the circle lies inside the exclusion region entirely (i.e., $k_*<K/2$), no attenuation arises.
    (b) Normalized sizes $K_T/K$ of the perfect transparency interval [\eqref{eq:trans-regime-TM}] for both layered and transversely isotropic stealthy hyperuniform media as a function of contrast ratio $\varepsilon_p/\varepsilon_q$.
    We consider five values of $\phi_p = \delta^+, 0.1,0.5,0.9,1-\delta^+$, where $\delta^+$ denotes an infinitesimally small positive number.  
    \label{fig:schm-transparency}}
\end{figure}

\section{Perfect Transparency Intervals of Stealthy Hyperuniform Systems}
\label{sec:trans}

We have previously shown that stealthy hyperuniform two-phase composites with $\spD{k}=0$ for $k<K$ can be perfectly transparent or, equivalently, have a zero imaginary part of effective dielectric constant, at the two-point level in a finite range of wave numbers for TE polarization in transversely isotropic media \cite{torquato_nonlocal_2021} and transverse polarization in 3D fully isotropic media \cite{torquato_nonlocal_2021} and layered media \cite{kim_effective_2023}.
Across space dimensions, the predictions of such perfect transparency intervals  have a similar form: 
\begin{align}
    0 \leq k_q < K_T \equiv \frac{K}{2 \sqrt{\varepsilon_* /\varepsilon_q}} ,    \label{eq:trans-regime-TM}
\end{align}
where the dielectric constant of the optimal reference phase $\varepsilon_*$ is given as
    \begin{numcases}{\varepsilon_*=}
        \E{\varepsilon},    \qquad ~~ \text{Transverse polarization in layered media}  \label{eq:eps_layered}\\
        \varepsilon_{BG}^{(2D)}, \qquad   \text{TE polarization in transversely isotropic media} \label{eq:eps_2D_TE} \\
        \varepsilon_{BG}^{(3D)},  \qquad  \text{Transverse polarizaion in 3D fully isotropic media}   ,\label{eq:eps_3D}
    \end{numcases}
where $\varepsilon_{BG}^{(dD)}$ is the Bruggeman (or self-consistent) approximation \cite{bruggeman_berechnung_1935, torquato_random_2002} of the effective static dielectric constant of $d$-dimensional two-phase composites \footnote{For a contrast ratio $\varepsilon_2/\varepsilon_1 \leq 4$, the Bruggeman formulas are very close to the Maxwell-Garnett formulas employed in Ref. \cite{torquato_nonlocal_2021}. The former offers a slightly better renormalization, since it uses the fact that the mean of the depolarization tensors is exactly zero; see Eq. (S105) of \supp. }.
The prediction of \eqref{eq:trans-regime-TM} in conjunction with \eqref{eq:eps_layered} leads to the predicted transparency condition, \eqref{eq:interval}, from the imaginary part of the two-point formula of the effective dielectric constant [\eqref{eq:im-layered-2pt}] for layered stealthy hyperuniform media.
Here, we show that the interval in the form of \eqref{eq:trans-regime-TM} still applies to TM polarization in transversely isotropic stealthy hyperuniform media at the two- and three-point levels:
\begin{align}
    \varepsilon_* = \E{\varepsilon},   \qquad &\text{TM polarizaion in transversely isotropic media }.  \label{eq:eps_2D_TM}
\end{align}
Moreover, we also show that the interval [\eqref{eq:trans-regime-TM}] for layered media still applies to the three-point level.

We now prove the perfect transparency interval for TM polarization in transversely isotropic media [\eqref{eq:trans-regime-TM} with \eqref{eq:eps_2D_TM}] at the two-point level.
Using the second-order formula of $\ETM{k_q}$ in \eqref{eq:eps-eff-2D_TM} and the nonlocal attenuation function [\eqref{eq:F-trans-Fourier}], one obtains
\begin{align}   \label{eq:trans-im-2pt}
    \Im[\ETM{k_q}] 
    \propto (\varepsilon_p-\varepsilon_q)^2 \Im[\ATM{2}{k_*, \E{\varepsilon}}]
    \propto 
    k_* \int_{\abs{k_*\uvect{y}+\vect{q}}=k_*} \spD{q} \dd{\vect{q}},
\end{align}
where $k_* = k_q \sqrt{\E{\varepsilon}/\varepsilon_q}$.
This relation implies that the effective attenuation at the two-point level comes solely from the on-shell scattering \cite{sheppard_greenfunction_2014,ong_control_2023}, which is proportional to an integral involving single-scattering contribution $\spD{q}$ for all possible scattering orientations, depicted as colored circles in Fig. \ref{fig:schm-transparency}(a).
Therefore, a stealthy hyperuniform system can be perfectly transparent if such a circle entirely lies inside the exclusion region (shown in a white disk) with $0\leq k_* = k_q \sqrt{\E{\varepsilon}/\varepsilon_q} <K/2$, which is identical to \eqref{eq:trans-regime-TM} in conjunction with \eqref{eq:eps_2D_TM}.
One can easily confirm the interval for TE polarization [\eqref{eq:trans-regime-TM} with \eqref{eq:eps_2D_TE}] because \eqref{eq:trans-im-2pt} applies to $\ETE{k_q}$ if we take $k_* = k_q \sqrt{\varepsilon_{BG}^{(2D)}/\varepsilon_q}$.

In the present work, we also sketch a proof that the perfect transparency interval [\eqref{eq:trans-regime-TM}] for layered media is still valid at the three-point level.
Accounting for the perfect transparency at the two-point level [see \eqref{eq:im-layered-2pt}], from the third-order formula of $\fn{\varepsilon_e^\perp}{k_q}$ [\eqref{eq:eps-eff-strat_perp}], we obtain 
$    \Im[\fn{\varepsilon_e^\perp}{k_q}]
    \propto (\varepsilon_p-\varepsilon_q)^3 \Im[\fn{A_3^\perp}{k_*, \E{\varepsilon}}]
$
for $0\leq k_*(= k_q \sqrt{\E{\varepsilon}/\varepsilon_q}) < K/2 $.
Thus, it is sufficient to show that 
\begin{align}   \label{eq:im-layered-3pt}
    \Im[\fn{A_3^\perp}{k_*, \E{\varepsilon}}] = 0,  \quad\text{for } 0\leq k_*<K/2. 
\end{align}
We outline the proof of \eqref{eq:im-layered-3pt} here (see Ref. \cite{kim_extraordinary_2023} and Sec. S2.D of \supp ~ for details).
We first decompose $\fn{A_3^\perp}{k_*;\E{\varepsilon}}$ into three terms as 
\begin{align}   \label{eq:A3-layered-PT}
    \fn{A_3^\perp}{k_*;\E{\varepsilon}}
        =&
        \frac{-1}{\phi_p}\frac{{k_*}^4}{(2\pi \E{\varepsilon})^2} 
        \Bigg\{ \fn{\CL{1}}{k_*} - {\phi_p}^2 (2\pi) \fn{\CL{2}}{k_*} - {\phi}_p \fn{\CL{3}}{k_*} 
        \Bigg\}, 
\end{align}
where $\fn{\CL{3}}{k}$ depends on the three-point statistics, but $\fn{\CL{1}}{k}$ and $\fn{\CL{2}}{k}$ do not:
\begin{align}
    \fn{\CL{1}}{k} 
    \equiv &  
    \qty[\frac{2\pi}{{k}^2} \F{1}{k}]^2
    \label{eq:C1_layered},  \\
    \fn{\CL{2}}{k}
    \equiv &
    \int_{-\infty}^\infty \dd{q_{1}} \qty[\frac{1}{(k+q_{1})^2-{k}^2}]^2 \spD{q_{1}} 
    \label{eq:C2_layered}, \\
    \fn{\CL{3}}{k}
    \equiv & 
    \int_{-\infty}^\infty \dd{q_{1z}} \int_{-\infty}^\infty \dd{q_{2}} \frac{1}{(k+q_{1})^2-{k}^2} \frac{1}{(k+q_{2})^2-{k}^2} 
        \E{\FTJ{p}{q_{1}} \FTJ{p}{-q_{1}+q_{2}} \FTJ{p}{-q_{2}}}
        \label{eq:C3_layered},
\end{align}
and $\FTJ{p}{k}$ denotes the Fourier transform of $\J{p}{x}$.
One can easily see that $\Im[\fn{\CL{1}}{k}] = 0$ for $0\leq k<K/2$ because $\Im[\fn{\CL{1}}{k}]
\propto \Im[\F{1}{k}] \propto \spD{0} + \spD{2k}$.
For $k<K/2$, the integrand in \eqref{eq:C2_layered} is a real-valued function that has a nonnegative and finite value at any $q_1$ and decays rapidly for large $\abs{q_1}$, implying that its integral $\fn{\CL{2}}{k}$ is also real-valued.
It is nontrivial to show that $\Im[\fn{\CL{3}}{k}]=0$ for $0\leq k<K/2$.
Using the stealthy hyperuniform condition [i.e., $\FTJ{p}{k}=0$ for $0\leq k<K$] and the large-$k$ scaling of $\FTJ{p}{k}$, we show that for $k<K/2$, the integrand in \eqref{eq:C3_layered} always has a finite value and decays rapidly for large $\abs{q_1}$, and thus its absolute value is integrable.
This implies that $\fn{\CL{3}}{k}$ is independent of the integration order.
Then, by using the following property of the Fourier transform, $\FTJ{p}{k}^* = \FTJ{p}{-k}$, and changing the integration order, we show that $\Im[\fn{\CL{3}}{k}]=0$ for $0\leq k<K/2$.
Thus, we prove the sufficient condition, \eqref{eq:im-layered-3pt}, of the perfect transparency interval at the three-point level.

 We also sketch a proof that the perfect transparency interval for TM polarization in transversely isotropic media [\eqref{eq:trans-regime-TM} with \eqref{eq:eps_2D_TM}] is still valid through the third-order terms.
Since the formula of $\ETM{k_q}$ [\eqref{eq:eps-eff-2D_TM}] for transversely isotropic media has the form identical to \eqref{eq:eps-eff-strat_perp} for layered media, it is sufficient to show that 
\begin{align}   \label{eq:im-TM-3pt}
    \Im[\ATM{3}{k_*,\E{\varepsilon}}] = 0,  \quad\text{for } 0\leq k_*<K/2. 
\end{align}
Its proof is analogous to the proof of \eqref{eq:im-layered-3pt} for layered media.
Thus, we similarly decompose $\ATM{3}{k_*, \E{\varepsilon}}$ into three terms as 
\begin{align*}
    \ATM{3}{k_*,\E{\varepsilon}}
        =&
        \frac{-1}{\phi_p}\frac{{k_*}^4}{(2\pi)^4 {\E{\varepsilon}}^2} 
        \Bigg\{ \fn{\CTM{1}}{k_*} - {\phi_p}^2 (2\pi)^2 \fn{\CTM{2}}{k_*} - {\phi}_p \fn{\CTM{3}}{k_*} 
        \Bigg\},
\end{align*}
where $\fn{\CTM{m}}{k}$ is the 2D counterpart of $\fn{\CL{m}}{k}$ given in \eqref{eq:C1_layered}-\eqref{eq:C3_layered}.
Using the general properties of $\FTJ{p}{k}$ for $d=2$ and the stealthy hyperuniformity condition, we can show that these three functions $\fn{\CTM{1}}{k}$, $\fn{\CTM{2}}{k}$, and $\fn{\CTM{3}}{k}$ are real-valued for $k<K/2$, and thus \eqref{eq:im-TM-3pt} is true.
We provide a detailed proof of \eqref{eq:im-TM-3pt} in Ref. \cite{kim_extraordinary_2023} and Sec. S3.D of \supp.

We now discuss the implications of the finite perfect transparency interval (trivially implying no Anderson localization \cite{anderson_absence_1958,mcgurn_anderson_1993, sheng_introduction_2006,aegerter_coherent_2009, izrailev_anomalous_2012,wiersma_disordered_2013}, in principle) for stealthy hyperuniform media across dimensions predicted in \eqref{eq:trans-regime-TM}.
As noted in the Introduction, the rapid convergence properties of strong-contrast expansions yield second-order truncations that already have high predictive power, as validated via FDTD simulations discussed in section \ref{sec:results} and in Ref. \cite{kim_effective_2023}. 
These outcomes imply that third- and higher-order contributions are negligibly small for relatively large contrast ratios ($\varepsilon_2/\varepsilon_1 \lesssim 10$), i.e., $\sum_{n=4}^\infty (\varepsilon_p-\varepsilon_q)^nA_n \approx 0$.
Therefore, the localization length associated with only possibly negligibly small higher-order contributions should be very large compared to any practically large sample size, and thus, there can be no Anderson localization in practice within the predicted perfect transparency interval in 1D and 2D stealthy hyperuniform media.
Indeed, for both layered and transversely isotropic stealthy hyperuniform media, our prediction of the finite perfect transparency interval [\eqref{eq:trans-regime-TM}] is even stronger, since we have shown they are exact through third-order terms.
The prediction for layered media is remarkable because the traditional understanding is that localization must occur for any type of disorder in one dimension \cite{sheng_introduction_2006, aegerter_coherent_2009, izrailev_anomalous_2012, wiersma_disordered_2013, yu_engineered_2021}.

We now show how the size $K_T$ of perfect transparency interval given in \eqref{eq:trans-regime-TM} varies with a contrast ratio $\varepsilon_p/\varepsilon_q$ and the volume fraction $\phi_p$ of phase with the higher dielectric constant.
In Fig. \ref{fig:schm-transparency}(b), we see such dependence of the size $K_T$ for layered and transversely isotropic stealthy hyperuniform media, given by \eqref{eq:trans-regime-TM} in conjunction with \eqref{eq:eps_layered} or, equivalently, \eqref{eq:eps_2D_TM}.
The size $K_T$ decreases with the contrast ratio $\varepsilon_p/\varepsilon_q$ because a higher contrast ratio strengthens scattering.
Similarly, as $\phi_p$ increases without changing the exclusion-region radius $K$, $K_T$ also decreases because $\phi_p$ is proportional to the density of scatterers or the frequency of scattering events.
One can also observe such a tendency that $K_T$ decreases with $\varepsilon_p/\varepsilon_q$ and $\phi_p$ in TE polarization in transversely isotropic media and 3D fully isotropic media because $\varepsilon_*$ given in \eqref{eq:eps_layered}-\eqref{eq:eps_2D_TM}  commonly increase with $\varepsilon_p/\varepsilon_q$.
For fixed values of $\phi_p$, $\varepsilon_p/\varepsilon_q$, and $K$, the size $K_T$ is the smallest for TM polarization in transversely isotropic media (and layered media), followed by 3D fully isotropic ones, and is the largest for TE polarization in transversely isotropic media. 
This ranking is also inversely related to the strength of reflectance (i.e., scattering) on an interface between two dielectric materials depending on polarizations, governed by the Fresnel equations \cite{jackson_classical_1999}.
For example, in the transversely isotropic media, TM-polarized waves are always reflected to a greater degree than TE-polarized waves.
We note that \eqref{eq:trans-regime-TM} is no longer valid at three trivial limits (i.e., $\varepsilon_p/\varepsilon_q=1$, $\phi_p=0$, and $\phi_p=1$) since the medium becomes homogeneous and perfectly transparent at any wave number $k_q$.

\section{Model Microstructures}
\label{sec:models}

We consider two models of 1D and 2D disordered hyperuniform media that enable us to generate 3D model microstructures of layered and transversely isotropic media, respectively.
For each symmetry, these models include nonstealthy
hyperuniform polydisperse sphere packings in a matrix and stealthy hyperuniform sphere packings in a matrix.
Since these models are particulate composites, we henceforth take the characteristic inhomogeneity length scale $\xi$ to be the mean particle separation $\rho^{-1/d}$, which is of the order of the mean nearest-neighbor distance $\ell_P$, (i.e., $\xi = \rho^{-1/d} \sim \ell_P$).
The particles of dielectric constant $\varepsilon_2$ are distributed throughout a matrix of dielectric constant $\varepsilon_1$.
We take the particle volume fraction $\phi_2 =0.2, 0.25$, respectively, in one and two dimensions.

\begin{figure}[h]
    \centering
    \includegraphics[width=0.9\textwidth]{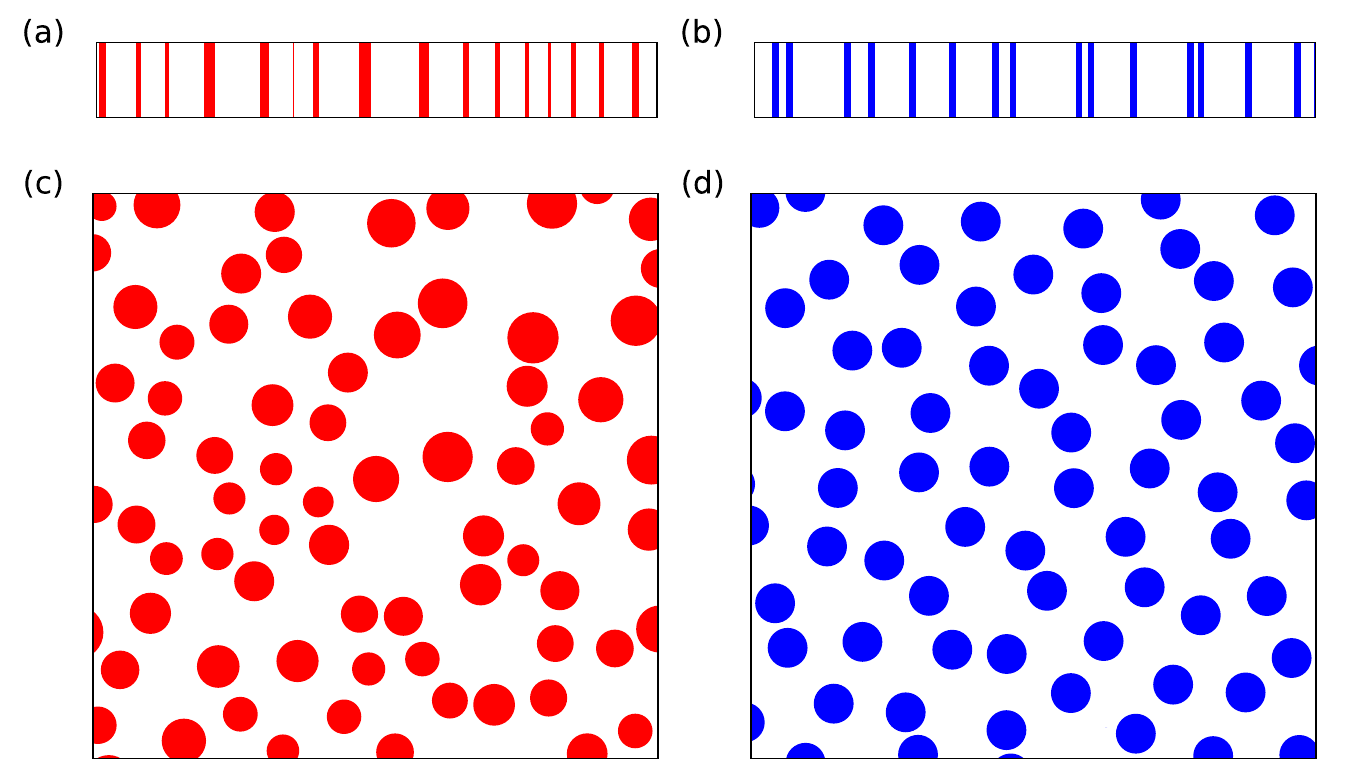}


    \caption{
    Top row shows representative images of configurations of two models of layered media with $\phi_2=0.2$: nonstealthy hyperuniform polydisperse sphere packings in a matrix (a) and stealthy hyperuniform sphere packings in a matrix (b).
    Bottom row shows representative images of configurations of two models of transversely isotropic media with $\phi_2=0.25$: nonstealthy hyperuniform polydisperse sphere packings in a matrix (c) and stealthy hyperuniform sphere packings in a matrix (d).
    \label{fig:model}}
\end{figure}

We numerically generate the configurations of these models (see Sec. S6 of \supp).
Figure \ref{fig:model} shows representative images of these configurations of all 1D and 2D models considered here.
For all models, we compute the spectral density (see Fig. \ref{fig:spd}), as described in section 5\ref{sec:chi}, and then use them to evaluate their nonlocal attenuation functions (see Fig. \ref{fig:F}).
We employ the configurations of hyperuniform models in FDTD simulations to extract their effective dynamic dielectric constants (see section \ref{sec:results}).


\subsection{Nonstealthy Hyperuniform Polydisperse Sphere Packings in a Matrix}
\label{sec:poly}

Nonstealthy hyperuniform (NSHU) packings of spheres with a polydispersity in size in a matrix can be constructed from nonhyperuniform progenitor point patterns via a tessellation-based procedure \cite{kim_new_2019, kim_methodology_2019}.
Specifically, the progenitor point patterns are the centers of 1D and 2D equilibrium packings of identical spheres with packing fraction $\phi_b$.
One begins with the Voronoi tessellation \cite{torquato_random_2002} of these progenitor point patterns.
We then move the particle center in a Voronoi cell to its centroid and rescale the particle such that the packing fraction inside this cell is identical to a prescribed value $\phi_2<1$.
The same process is repeated over all cells.
The final packing fraction is $\phi_2 = \sum_{j=1}^N \fn{v_1}{a_j}/V_\mathfrak{F} = \rho \fn{v_1}{a}$, where $\rho$ is the number density of particle centers and $a$ represents the mean sphere radius.
In the thermodynamic limit, the spectral densities of the resulting NSHU media exhibit a power-law scaling $\spD{\vect{k}} \sim \abs{\vect{k}}^4$ for small wave numbers \cite{kim_methodology_2019}, and thus these systems are of class I.
\newline

\subsection{Stealthy Hyperuniform Sphere Packings in a Matrix}\label{sec:stealthy}

Stealthy hyperuniform (SHU) two-phase media have $\spD{\vect{k}}=0$ for the finite range $0<\abs{\vect{k}}\leq K$, called the {\it exclusion region}.
We specifically consider $d$-dimensional SHU sphere packings in a matrix with packing fraction $\phi_2$.
For such SHU media, the degree of stealthiness $\chi$ is measured by the ratio of the number of the wave vectors within the exclusion region in the Fourier space to the total degrees of freedom, i.e., $\chi = K / (2\pi \rho)$ in one dimension, and $\chi=K^2 / (16\pi \rho)$ in two dimensions.
These SHU systems are highly degenerate and disordered if $\chi <1/3$ in one dimension or $\chi<1/2$ in two and three dimensions \cite{zhang_ground_2015}.
Thus, we consider 1D and 2D cases for $\chi=0.3$ in the present work.

We numerically generate such 1D and 2D SHU media in the following two-step procedure. 
First, we generate point configurations consisting of $N$ particles in a fundamental cell $\mathfrak{F}$ under periodic boundary conditions via the collective-coordinate optimization technique \cite{uche_constraints_2004, batten_classical_2008, zhang_ground_2015}, which finds numerically the ground-state configurations of the following potential energy;
\begin{equation*}\label{eq:CC_potential}
\fn{\Phi}{\vect{r}^N} =\frac{1}{V_\mathfrak{F}} \sum_{\vect{k}} \fn{\tilde{v}}{\vect{k}}\fn{S}{\vect{k}} +  \sum_{i <j} \fn{u}{r_{ij}},
\end{equation*}
where $V_\mathfrak{F}$ is the volume of $\mathfrak{F}$, $\fn{\tilde{v}}{\vect{k}}=\fn{\Theta}{K-\abs{\vect{k}}}$, $\Theta(x)$ (equal to 1 for $x>0$ and zero otherwise) is the Heaviside step function, $\fn{u}{r}=(1-r/\sigma)^2\fn{\Theta}{\sigma-r}$ \cite{zhang_can_2017}. 
The consequent ground-state configurations, if they exist, are still disordered, stealthy, and hyperuniform, and their nearest-neighbor distances are larger than the length scale $\sigma$.
Importantly, without the soft-core repulsion $\fn{u}{r}$, one cannot generate disordered SHU packings in a matrix with $\phi_2\geq0.2$ for $d=1,2$ \cite{zhang_ground_2015}.
Finally, to create SHU media, we follow Ref. \cite{zhang_transport_2016} by circumscribing the points by identical spheres of radius $a<\sigma/2$ under the constraint that they cannot overlap.

\subsection{Spectral Densities for the Hyperuniform Models}    \label{sec:chi}

\begin{figure}[h]
    \includegraphics[width=0.49\textwidth]{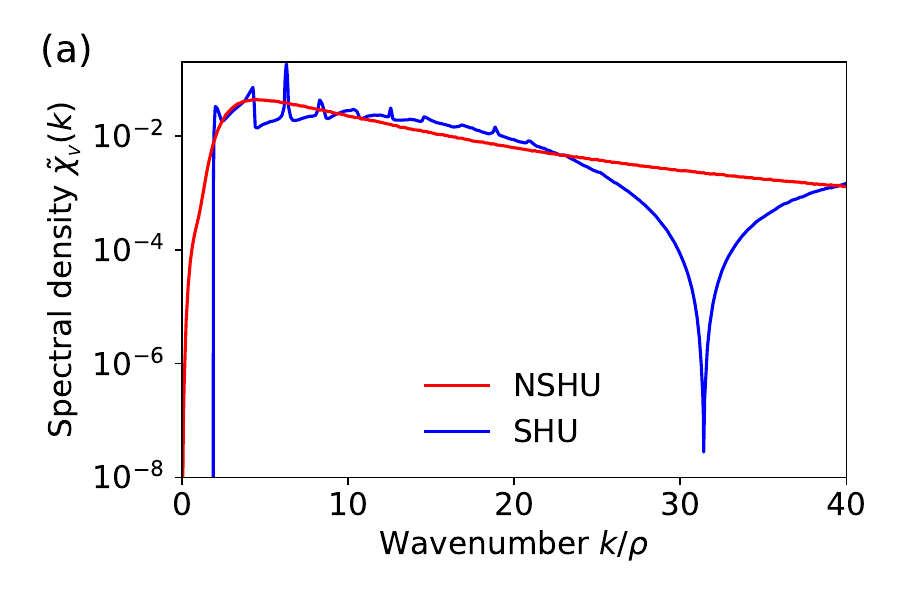}
    \includegraphics[width=0.49\textwidth]{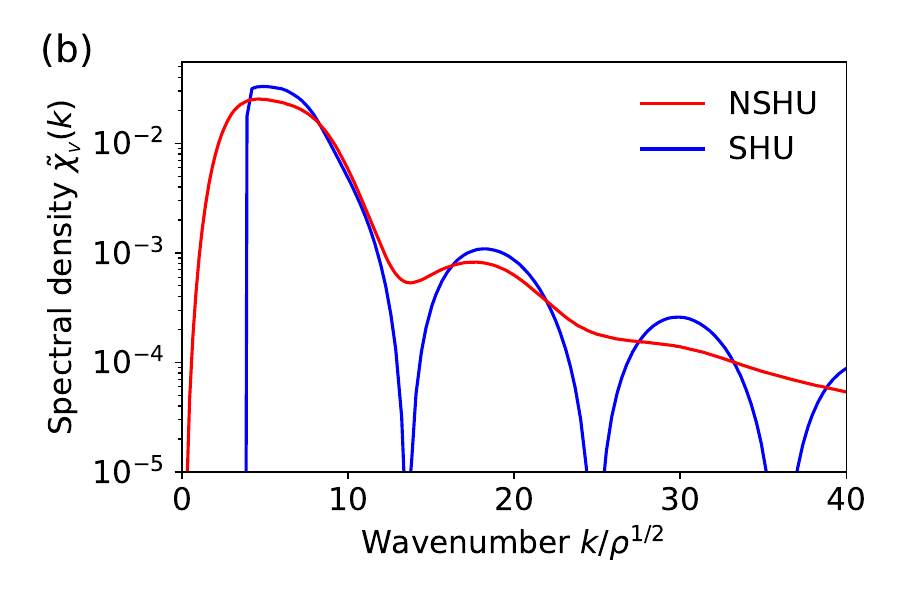}
    \caption{ Semi-log plots of spectral densities $\spD{k}$ as functions of dimensionless wave number for the two hyperuniform models in (a) one and (b) two dimensions: NSHU and SHU models.
    The particle-phase volume fractions are $\phi_2=0.2$ in (a) and $\phi_2=0.25$ in (b). \label{fig:spd}}
\end{figure}

Here, we plot the radial spectral density $\spD{k}$ of 1D and 2D hyperuniform models; see Fig. \ref{fig:spd}.
We take the particle-phase volume fraction $\phi_2=0.2$ and $\phi_2=0.25$ for 1D and 2D cases, respectively.
For all models, we numerically obtain $\spD{k}$ by using \eqref{eq:spd-packing} from numerically generated configurations.

Across all length scales, the spectral densities of these two models are considerably different from one another.
In the long-wavelength regime ($k/\rho^{1/d} \ll 1$), SHU media suppress volume-fraction fluctuations [i.e., $\spD{k}=0$] to a greater degree than the NSHU media over a wider range of wavelengths.
In the small-wavelength regime ($k/\rho^{1/d} \gg 10$), the oscillations of $\spD{k}$ reflect the particle-size distribution.
Specifically, the spectral densities of the SHU model exhibit strong oscillations because this model consists of particles of identical size, whereas those of the NSHU model exhibit weaker oscillations because of a broad disparity in particle sizes.

\begin{figure}[h]
  {\includegraphics[width=0.49\textwidth]{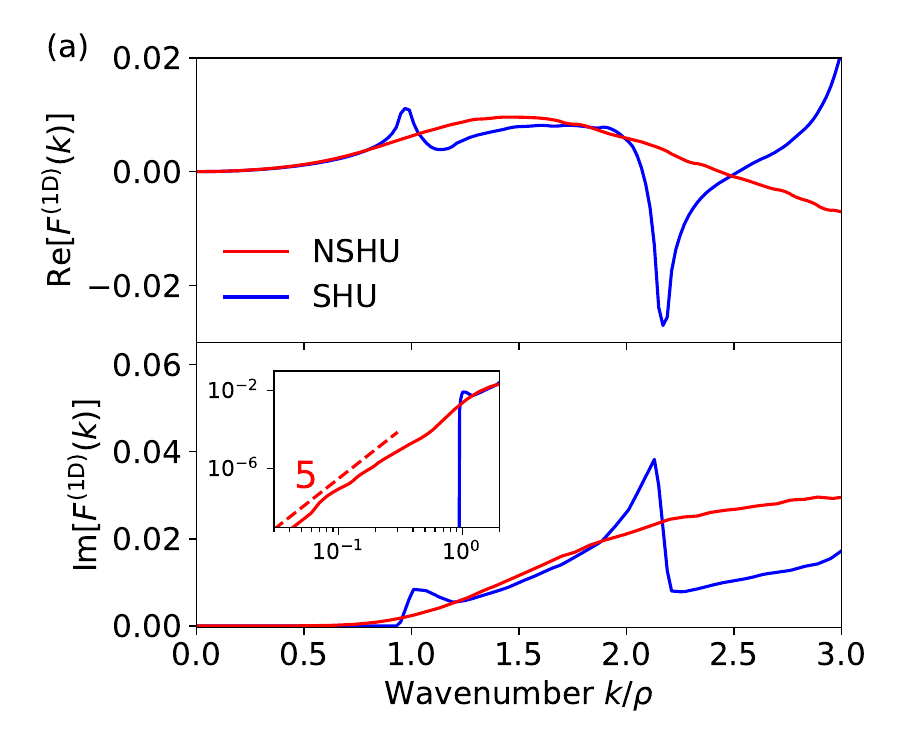}}
  {\includegraphics[width=0.49\textwidth]
  {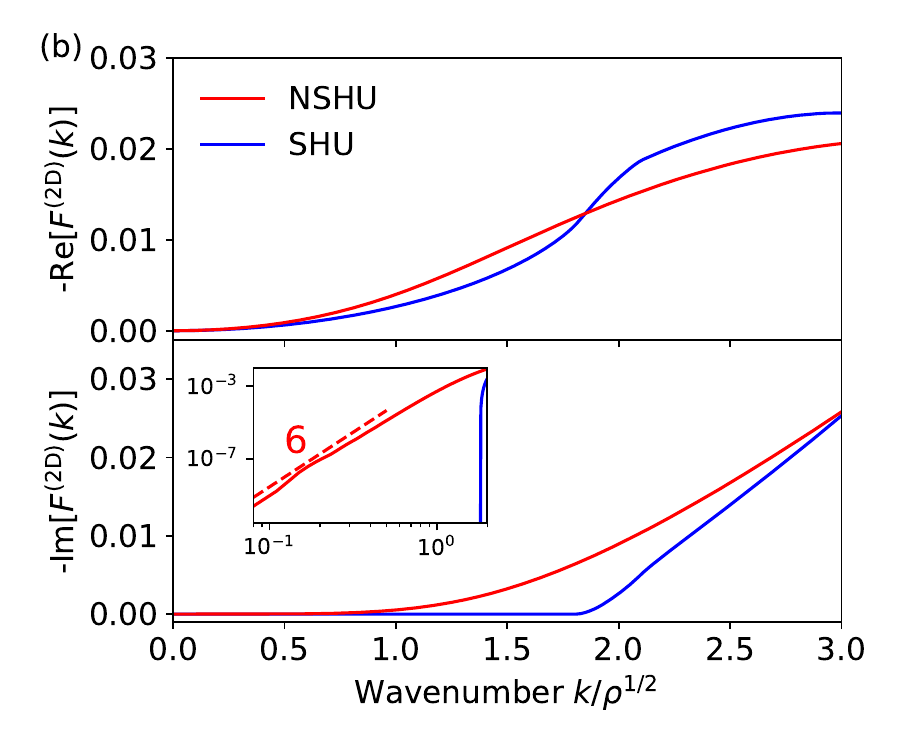}}
  \caption{Real (upper) and imaginary (lower) parts of the nonlocal attenuation functions versus dimensionless wave number for the two models in (a) one and (b) two dimensions: NSHU and SHU media.
  The packing fractions in (a) and (b) are $\phi_2=0.2$ and $0.25$, respectively.
  Values in (a) are computed from the spectral densities in Fig. \ref{fig:spd}(a) and \eqref{eq:F-strat-Fourier}.
  In (b), negative values of $\F{2}{k}$ are computed from the spectral densities in Fig. \ref{fig:spd}(b) and \eqref{eq:F-trans-Fourier}.
  In the lower panels, each inset is in a log-log scale, where the dashed lines with numbers depict the small-$k$ scalings for NSHU systems predicted in \eqref{eq:asymptotic-layered} and \eqref{eq:asymptotic-trans-TM}.  
  \label{fig:F}}
\end{figure}

\begin{figure}[H]
          \includegraphics[width=0.49\textwidth]{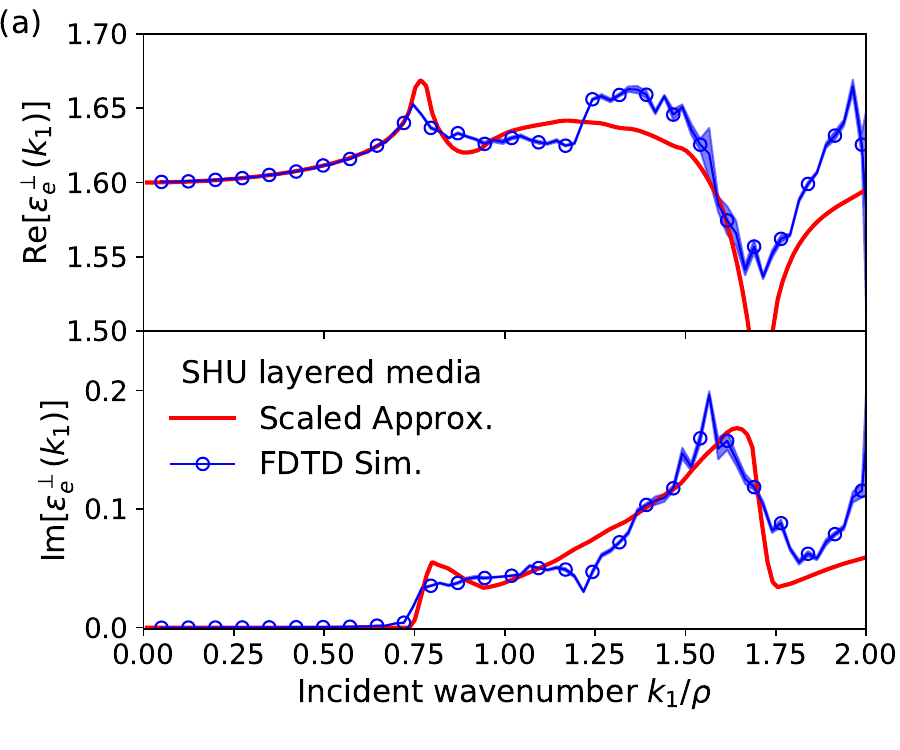}
          \includegraphics[width=0.49\textwidth]{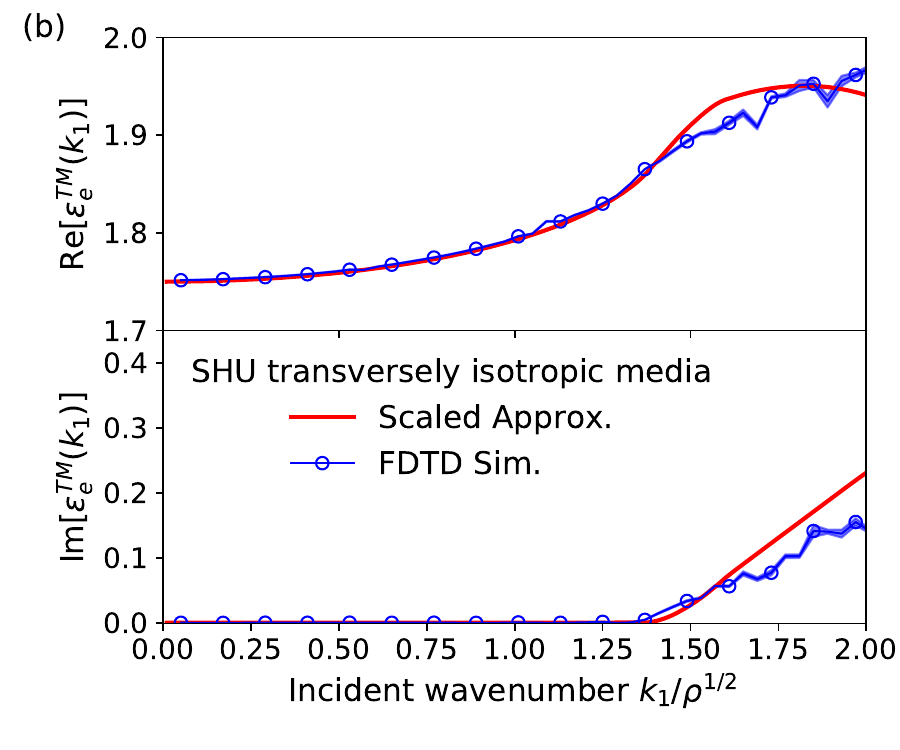}
      \caption{
      Comparison of the predictions of the second-order strong-contrast approximations for the effective dynamic dielectric constant as a function of the dimensionless wave number for SHU systems with various symmetries to FDTD simulations.
      We consider (a) layered media with $\phi_2=0.20$,  $\chi=0.30$, and $\varepsilon_2/\varepsilon_1=4$, and (b) transversely isotropic media with $\phi_2=0.25$, $\chi=0.30$, and $\varepsilon_2/\varepsilon_1=4$.
      Values of approximations in (a) and (b) are computed from \eqref{eq:eps-eff-strat_perp} and \eqref{eq:eps-eff-2D_TM} with $A_3(k)=0$, respectively.
      \label{fig:sim}}
  \end{figure}

\section{Results}   \label{sec:results}

We now show how our multiple-scattering approximations for layered media [\eqref{eq:eps-eff-strat_perp}] and transversely isotropic media [\eqref{eq:eps-eff-2D_TM}] enable us to accurately predict the real and imaginary parts of the effective dynamic dielectric constants for anisotropic disordered two-phase media.
For simplicity, we consider the second-order approximations by neglecting the third-order terms $A_3^\perp$ and $A_3^{TM}$ in \eqref{eq:eps-eff-strat_perp} and \eqref{eq:eps-eff-2D_TM}, respectively.
These formulas depend on the nonlocal attenuation functions that incorporate microstructural information via the spectral density $\spD{k}$ only.
We begin by computing the nonlocal attenuation functions $\F{1}{k}$ [given in \eqref{eq:F-strat-Fourier}] and $\F{2}{k}$ [given in \eqref{eq:F-trans-Fourier}] of the two hyperuniform models from $\spD{k}$; see Fig. \ref{fig:F}.
For SHU media, the imaginary parts $\Im[\F{1}{k}]$ and $\Im[\F{2}{k}]$ are exactly zero up to a finite wave number $k$, leading to the prediction of perfect transparency interval; see \eqref{eq:trans-regime-TM}.
By contrast, for the NSHU media, the imaginary parts are close but not equal to zero for small wave numbers, as shown in the insets of Fig. \ref{fig:F}.

To validate the accuracy of our two second-order approximations, we show that the predicted effective dielectric constants are in excellent agreement with those extracted from FDTD simulations.
We make such comparisons in Fig. \ref{fig:sim} for the disordered SHU media.
(We present analogous comparisons for the NSHU media in Sec. S7 of \supp.)
This model can provide stringent tests of the predictive power of the approximations at finite wave numbers because they are characterized by nontrivial spatial correlations at intermediate length scales.
For both layered media and transversely isotropic ones, the real and imaginary parts of these predictions show excellent agreement with the results from FDTD simulations up to $k_1 \rho^{-1/d} \lesssim 1.5$, which is well beyond the long-wavelength regime.
It is noteworthy that our second-order  approximation formulas [\eqref{eq:eps-eff-strat_perp} and \eqref{eq:eps-eff-2D_TM}] can be used to accurately estimate the effective dielectric response for higher contrast ratios (e.g., $9\lesssim \varepsilon_2/\varepsilon_1\lesssim 12$); see, for example, the predicted transparency interval as a function of contrast ratio $\varepsilon_2/\varepsilon_1$ as shown in Fig. \ref{fig:schm-transparency}(b).
For the imaginary parts, both approximations and simulations consistently show that $\Im[\fn{\varepsilon_e}{k_1}]=0$ within the finite perfect transparency intervals predicted in \eqref{eq:trans-regime-TM}.
We also observe that the real parts $\Re[\fn{\varepsilon_e}{k_1}]$ increase with $k_1$ (i.e., normal dispersion) within the perfect transparency intervals and then decrease with $k_1$ (i.e., anomalous dispersion) outside those intervals.
Such a spectral dependence of the real parts comes from the fact that our two second-order strong-contrast formulas satisfy the Kramers-Kronig relations \cite{jackson_classical_1999}, as shown in Refs. \cite{torquato_nonlocal_2021, kim_effective_2023}.
We analytically and numerically show that our second-order formulas for both layered and transversely isotropic media meet the Kramers-Kronig relations in Sec. S8 of \supp.

\begin{figure}[h]
    {
        \includegraphics[width=0.49\textwidth]{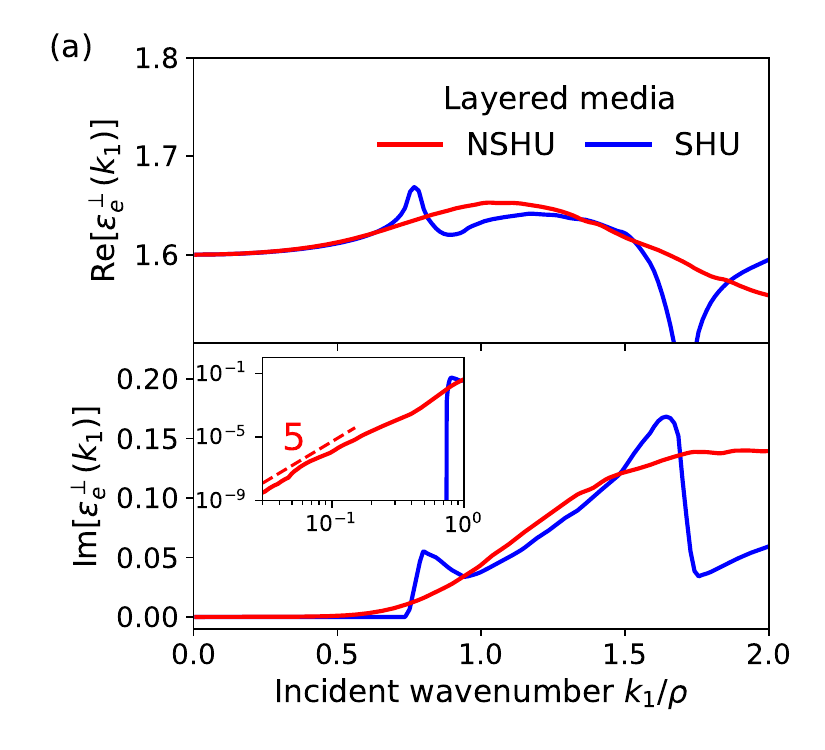}
    }
    {
        \includegraphics[width=0.49\textwidth]{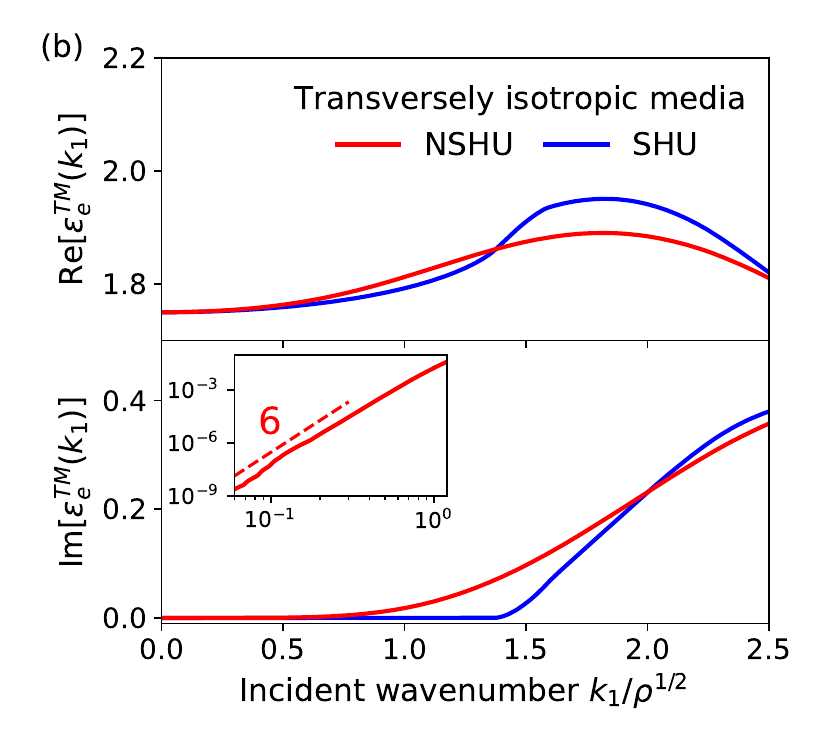}
    }
    \caption{ 
    Predictions of the scaled second-order strong-contrast approximations for the effective dynamic dielectric constant as a function of the dimensionless incident wave number for the two hyperuniform models of (a) layered media and (b) transversely isotropic media with a contrast ratio $\varepsilon_2/\varepsilon_1 = 4$: SHU and NSHU media.
    We consider the packing fractions $\phi_2=0.2$ and $0.25$ in (a) and (b), respectively.
    Values in (a) and (b) are computed from \eqref{eq:eps-eff-strat_perp} and \eqref{eq:eps-eff-2D_TM}, respectively, at the two-point level.
    In the lower panels, each inset is in a log-log scale, where the dashed lines with numbers depict the small-$k$ scaling for the NSHU systems, predicted in \eqref{eq:asymptotic-layered} and \eqref{eq:asymptotic-trans-TM}.  
    \label{fig:pred}}
\end{figure}

Having established the accuracy of the two second-order strong-contrast approximations [\eqref{eq:eps-eff-strat_perp} and \eqref{eq:eps-eff-2D_TM}] for anisotropic disordered media in Fig. \ref{fig:sim} and Fig. S3 of \supp, we apply them to estimate $\fn{\varepsilon_e}{k_q}$ for models of disordered hyperuniform layered and transversely isotropic media: the NSHU and SHU models
Figure \ref{fig:pred} shows the predictions at a fixed contrast ratio $\varepsilon_2/\varepsilon_1=4$.
For these two disordered models of layered media and transversely isotropic media, the real parts of the effective dielectric constants $\Re[\varepsilon_e]$, associated with the effective wave speed, similarly increase with $k_1$ for small wave numbers ($k_1\rho^{-1/d}\lesssim 1.0$).
We see that the imaginary parts of the effective dielectric constants have distinctly different small-wave number behaviors of the two models of both layered and transversely isotropic media. 
The NSHU media exhibit $\Im[\varepsilon_e]\sim {k_1}^{4+d}$ in the limit $k_1\to 0$, as predicted in both \eqref{eq:asymptotic-layered} and \eqref{eq:asymptotic-trans-TM}.
While the imaginary part of such a NSHU medium is expected to be much smaller than that of typical nonhyperuniform ones, it cannot be exactly zero at any wave number, unlike a SHU medium, as discussed in section \ref{sec:trans}.

\section{Conclusion and Discussion}
\label{sec:conclusion}

We have demonstrated that the second- or third-order truncations of the rapidly converging strong-contrast expansion \cite{torquato_nonlocal_2021} of the effective dielectric constant tensor $\fn{\tens{\varepsilon}_e}{\vect{k}_q, \omega}$ provide a powerful theoretical tool to extract accurate multiple-scattering approximations suited for various microstructural symmetries.
This task was accomplished by judiciously tuning the tensor expansion parameter $\tens{L}_p^{(q)}$ (see Sec. S1 of \supp), which is determined by choosing the exclusion volume shape with an appropriate symmetry around the singularity of the dyadic Green's function \cite{yaghjian_electric_1980,torquato_nonlocal_2021,kim_effective_2023}.
We have shown that the third-order truncations of such series yield closed-form formulas applicable to layered media [\eqref{eq:eps-eff-strat_perp} and \eqref{eq:eps-eff-strat_z}] and to transversely isotropic media [\eqref{eq:eps-eff-2D_TM} and \eqref{eq:2pt-trans-TE}], respectively; see section \ref{sec:theory}.
These formulas incorporate microstructural information through two- and three-point correlation functions.
Including the formulas for 3D fully isotropic composites derived in Ref. \cite{torquato_nonlocal_2021}, we have obtained a family of multiple-scattering approximations applicable to three distinct types of microstructural symmetries.
In the present work, since \eqref{eq:eps-eff-strat_z} is purely static, and \eqref{eq:2pt-trans-TE} is similar to what was studied in Ref. \cite{torquato_nonlocal_2021}, we focused on two key formulas \eqref{eq:eps-eff-strat_perp} and \eqref{eq:eps-eff-2D_TM}, which are for transverse polarization in layered media and TM polarization in transversely isotropic media, respectively.

Using these formulas, we proved that the finite perfect transparency interval of SHU media, predicted in \eqref{eq:trans-regime-TM}, is exact through the third-order terms for both transversely isotropic media (TM case) and layered media.
The latter result further validates the accuracy of our previous prediction from the second-order approximation in Ref. \cite{kim_effective_2023}; see \eqref{eq:im-layered-2pt}.
The rapid convergence properties of strong-contrast expansions (see Sec. S5 of \supp) and the high predictive power of their second-order truncations validated via FDTD simulations imply that higher-order contributions are negligibly small.
This trivially implies that, in principle,  there can be no Anderson localization within the predicted perfect transparency interval in SHU media.
In practice, this means that the localization length, associated with only possibly negligibly small higher-order contributions, should be very large compared to any practically large sample size.
Indeed, for SHU layered and transversely isotropic media, our predictions of the perfect transparency interval \eqref{eq:trans-regime-TM} are even stronger, since we have shown they are exact through third-order terms, and thus, there can be no localization rigorously through the third-order terms in the thermodynamic limit when the contrast ratio is sufficiently small.
As noted in Ref. \cite{kim_effective_2023}, this prediction is remarkable because the traditional understanding is that localization must occur for any type of disorder in one dimension \cite{sheng_introduction_2006, aegerter_coherent_2009, izrailev_anomalous_2012, wiersma_disordered_2013, yu_engineered_2021}.
In future work, we will systematically study the localization length of 1D SHU media \cite{Kl_local_2023}.

We have applied our formulas to estimate $\fn{\varepsilon_e}{k_q}$ for two models of disordered hyperuniform layered and transversely isotropic media: NSHU and SHU models.
For all hyperuniform models considered here, we corroborated that the second-order formulas for layered media [$\eqref{eq:eps-eff-strat_perp}$, where $A_3^\perp=0$] and transversely isotropic media [\eqref{eq:eps-eff-2D_TM}, where $A_3^{TM}=0$] are already very accurate well beyond the long-wavelength regime (i.e., $k_1 \rho^{-1/d} \lesssim 1.5$) by showing excellent agreement with FDTD simulations; see section \ref{sec:results} and Sec. S7 of \supp.
The real parts $\Re[\varepsilon_e]$ of the effective dielectric constants exhibit similar behavior (i.e., $\Re[\varepsilon_e]$ increases with $k_1$) in the long-wavelength regime, implying the effective wave speed is largely insensitive to microstructures; see Fig. \ref{fig:pred}.
However, our predictions reveal that the two models exhibit distinctly different effective attenuation behaviors; see the insets of Fig. \ref{fig:pred}.
While NSHU models do not have a finite perfect transparency interval, they can exhibit a wide range of attenuation behaviors by tuning the exponent $\alpha$ in a power-law scaling of the spectral density, as predicted in \eqref{eq:asymptotic-layered} and \eqref{eq:asymptotic-trans-TM}.

Our findings also have important practical implications. 
For example, combining our accurate second-order formulas depending solely on the spectral density with the methods to construct two-phase media with a prescribed spectral density \cite{chen_designing_2018,shi_computational_2023,uche_constraints_2004,batten_classical_2008,zhang_ground_2015} enables one to take an inverse-design approach \cite{torquato_inverse_2009} to engineering anisotropic dielectric materials with novel wave properties.
Notably, such computationally designed composite media can be readily fabricated via vacuum deposition \cite{affinito_new_1996}, spin-coating \cite{lee_LayerbyLayer_2001}, 2D photolithographic \cite{zhao_assembly_2018}, and 3D printing techniques \cite{tumbleston_continuous_2015}.
As demonstrated in Ref. \cite{zhou_ultrabroadband_2020}, such anisotropic properties can be employed to design broadband polarizers that transmit waves up to selected wave numbers, depending on the polarization.
Thus, our results offer promising prospects for engineering novel optoelectronic materials and devices across space dimensions by engineering the spectral density of the systems \cite{chen_designing_2018}.

\section*{Appendix}
\appendix
\section{Simulations}   \label{sec:FDTD}

We validate the accuracy of our second-order formulas for the effective dielectric constants by showing very good agreement with full-waveform simulations \cite{taflove_advances_2013} via an open-source FDTD package MEEP \cite{oskooi_meep_2010} in both one and two dimensions.
We take the matrix to be the reference phase (i.e., $q=1$) and the particles to be the polarized phase (i.e., $p=2$) and set the phase contrast ratio as $\varepsilon_2/\varepsilon_1 = 4$.
We directly extract $\fn{\varepsilon_e}{k_1}$ [i.e., $\fn{\varepsilon_e^\perp }{k_1}$ for layered media or $\fn{\varepsilon_e^{TM} }{k_1}$ for transversely isotropic media] from the nonlocal constitutive relation $\fn{\varepsilon_e }{k_1} = \E{\fn{\tilde{D}}{k_e,\omega}}/\E{\fn{\tilde{E}}{k_e,\omega}}$ at a given frequency $\omega$, as was done in Refs. \cite{torquato_nonlocal_2021,kim_effective_2023}.
Here, $\E{\fn{\tilde{D}}{k_e,\omega}}$ and $\E{\fn{\tilde{E}}{k_e,\omega}}$ are the spatial Fourier transforms of the ensemble averages of dielectric displacement field $\E{\fn{D}{x,\omega}}$ and electric field $\E{\fn{E}{x,\omega}}$ at the complex-valued {\it effective wave number} $k_e$, respectively; see details in Sec. S6 of \supp. 
We also measure the transmittance spectra $T$ through the disordered two-phase media.

\begin{backmatter}
    \bmsection{Funding}
    Army Research Office (W911NF-22-2-0103)  
    
    \bmsection{Acknowledgments}
    Simulations were performed on computational resources managed and supported by the Princeton Institute for Computational Science and Engineering (PICSciE).
    We thank J. La, M. C. Rechtsman and J. Karcher for very helpful discussions.

    \bmsection{Disclosures}
    \noindent The authors declare no conflicts of interest.
    
    \bmsection{Data availability} Data underlying the results presented in this paper are not publicly available at this time but may be obtained from the authors upon reasonable request.
    
    \bmsection{Supplemental document} See Supplement 1 for supporting content.
    \end{backmatter}


\begin{thebibliography}{100}
    \newcommand{\enquote}[1]{``#1''}
    
    \bibitem{florescu_designer_2009}
    M.~Florescu, S.~Torquato, and P.~Steinhardt, \enquote{Designer disordered
      materials with large, complete photonic band gaps,}
      {\protect\JournalTitle{Proceedings of the National Academy of Sciences of the
      United States of America}} \textbf{106}, 20658--20663 (2009).
    
    \bibitem{izrailev_anomalous_2012}
    F.~M. Izrailev, A.~A. Krokhin, and N.~M. Makarov, \enquote{Anomalous
      localization in low-dimensional systems with correlated disorder,}
      {\protect\JournalTitle{Physics Reports}} \textbf{512}, 125--254 (2012).
    
    \bibitem{man_isotropic_2013}
    W.~Man, M.~Florescu, E.~Williamson, Y.~He, S.~Hashemizad, B.~Leung, D.~Liner,
      S.~Torquato, P.~Chaikin, and P.~Steinhardt, \enquote{Isotropic band gaps and
      freeform waveguides observed in hyperuniform disordered photonic solids,}
      {\protect\JournalTitle{Proc. Natl. Acad. Sci. U.S.A.}} \textbf{110},
      15886--15891 (2013).
    
    \bibitem{ma_3d_2016}
    T.~Ma, H.~Guerboukha, M.~Girard, A.~D. Squires, R.~A. Lewis, and
      M.~Skorobogatiy, \enquote{3d printed hollow-core terahertz optical waveguides
      with hyperuniform disordered dielectric reflectors,}
      {\protect\JournalTitle{Advanced Optical Materials}} \textbf{4}, 2085--2094
      (2016).
    
    \bibitem{leseur_highdensity_2016}
    O.~Leseur, R.~Pierrat, and R.~Carminati, \enquote{High-density hyperuniform
      materials can be transparent,} {\protect\JournalTitle{Optica}} \textbf{3},
      763--767 (2016).
    
    \bibitem{wu_effective_2017}
    B.~Wu, X.~Sheng, and Y.~Hao, \enquote{Effective media properties of
      hyperuniform disordered composite materials,} {\protect\JournalTitle{PLoS
      One}} \textbf{12}, e0185921 (2017).
    
    \bibitem{xu_microstructure_2017}
    Y.~Xu, S.~Chen, P.-E. Chen, W.~Xu, and Y.~Jiao, \enquote{Microstructure and
      mechanical properties of hyperuniform heterogeneous materials,}
      {\protect\JournalTitle{Physical Review E}} \textbf{96}, 043301 (2017).
    
    \bibitem{froufe-perez_band_2017}
    L.~{Froufe-P{\'e}rez}, M.~Engel, J.~S{\'a}enz, and F.~Scheffold, \enquote{Band
      gap formation and anderson localization in disordered photonic materials with
      structural correlations,} {\protect\JournalTitle{Proceedings of the National
      Academy of Sciences of the United States of America}} \textbf{114},
      9570--9574 (2017).
    
    \bibitem{gkantzounis_freeform_2017}
    G.~Gkantzounis and M.~Florescu, \enquote{Freeform phononic waveguides,}
      {\protect\JournalTitle{Crystals}} \textbf{7}, 353 (2017).
    
    \bibitem{chen_designing_2018}
    D.~Chen and S.~Torquato, \enquote{Designing disordered hyperuniform two-phase
      materials with novel physical properties,} {\protect\JournalTitle{Acta
      Mater.}} \textbf{142}, 152--161 (2018).
    
    \bibitem{gorsky_engineered_2019}
    S.~Gorsky, W.~A. Britton, Y.~Chen, J.~Montaner, A.~Lenef, M.~Raukas, and
      L.~Dal~Negro, \enquote{Engineered hyperuniformity for directional light
      extraction,} {\protect\JournalTitle{APL Photonics}} \textbf{4}, 110801
      (2019).
    
    \bibitem{kim_multifunctional_2020}
    J.~Kim and S.~Torquato, \enquote{Multifunctional composites for elastic and
      electromagnetic wave propagation,} {\protect\JournalTitle{Proc. Natl. Acad.
      Sci. U.S.A.}} \textbf{117}, 8764--8774 (2020).
    
    \bibitem{rohfritsch_impact_2020}
    A.~Rohfritsch, J.-M. Conoir, T.~{Valier-Brasier}, and R.~Marchiano,
      \enquote{Impact of particle size and multiple scattering on the propagation
      of waves in stealthy-hyperuniform media,} {\protect\JournalTitle{Phys. Rev.
      E}} \textbf{102}, 053001 (2020).
    
    \bibitem{yu_engineered_2021}
    S.~Yu, C.-W. Qiu, Y.~Chong, S.~Torquato, and N.~Park, \enquote{Engineered
      disorder in photonics,} {\protect\JournalTitle{Nat Rev Mater}} \textbf{6},
      226--243 (2021).
    
    \bibitem{romero-garcia_wave_2021}
    V.~{Romero-Garc{\'i}a}, {\'E}.~Ch{\'e}ron, S.~Kuznetsova, J.-P. Groby,
      S.~F{\'e}lix, V.~Pagneux, and L.~M. {Garcia-Raffi}, \enquote{Wave transport
      in 1d stealthy hyperuniform phononic materials made of non-resonant and
      resonant scatterers,} {\protect\JournalTitle{APL Mater.}} \textbf{9}, 101101
      (2021).
    
    \bibitem{vynck_light_2021}
    K.~Vynck, R.~Pierrat, R.~Carminati, L.~S. {Froufe-P{\'e}rez}, F.~Scheffold,
      R.~Sapienza, S.~Vignolini, and J.~J. S{\'a}enz, \enquote{Light in correlated
      disordered media,} {\protect\JournalTitle{ArXiv210613892 Cond-Mat
      Physicsphysics}}  (2021).
    
    \bibitem{kim_bragg_2022}
    D.~Kim and E.~J. Heller, \enquote{Bragg scattering from a random potential,}
      {\protect\JournalTitle{Phys. Rev. Lett.}} \textbf{128}, 200402 (2022).
    
    \bibitem{sgrignuoli_subdiffusive_2022}
    F.~Sgrignuoli, S.~Torquato, and L.~Dal~Negro, \enquote{Subdiffusive wave
      transport and weak localization transition in three-dimensional stealthy
      hyperuniform disordered systems,} {\protect\JournalTitle{Phys. Rev. B}}
      \textbf{105}, 064204 (2022).
    
    \bibitem{granchi_nearfield_2022}
    N.~Granchi, R.~Spalding, M.~Lodde, M.~Petruzzella, F.~W. Otten, A.~Fiore,
      F.~Intonti, R.~Sapienza, M.~Florescu, and M.~Gurioli, \enquote{Near-field
      investigation of luminescent hyperuniform disordered materials,}
      {\protect\JournalTitle{Adv. Opt. Mater.}} \textbf{10}, 2102565 (2022).
    
    \bibitem{tavakoli_65_2022}
    N.~Tavakoli, R.~Spalding, A.~Lambertz, P.~Koppejan, G.~Gkantzounis, C.~Wan,
      R.~R{\"o}hrich, E.~Kontoleta, A.~F. Koenderink, R.~Sapienza, M.~Florescu, and
      E.~{Alarcon-Llado}, \enquote{Over 65\% sunlight absorption in a 1 {$\mu$}m si
      slab with hyperuniform texture,} {\protect\JournalTitle{ACS Photonics}}
      \textbf{9}, 1206--1217 (2022).
    
    \bibitem{cheron_wave_2022}
    {\'E}.~Ch{\'e}ron, S.~F{\'e}lix, J.-P. Groby, V.~Pagneux, and
      V.~{Romero-Garc{\'i}a}, \enquote{Wave transport in stealth hyperuniform
      materials: The diffusive regime and beyond,} {\protect\JournalTitle{Appl.
      Phys. Lett.}} \textbf{121}, 061702 (2022).
    
    \bibitem{klatt_wave_2022}
    M.~A. Klatt, P.~J. Steinhardt, and S.~Torquato, \enquote{Wave propagation and
      band tails of two-dimensional disordered systems in the thermodynamic limit,}
      {\protect\JournalTitle{Proc. Natl. Acad. Sci.}} \textbf{119}, e2213633119
      (2022).
    
    \bibitem{tang_hyperuniform_2023}
    K.~Tang, Y.~Wang, S.~Wang, D.~Gao, H.~Li, X.~Liang, P.~Sebbah, J.~Zhang, and
      J.~Shi, \enquote{Hyperuniform disordered parametric loudspeaker array,}
      (2023).
    
    \bibitem{shi_computational_2023}
    W.~Shi, D.~Keeney, D.~Chen, Y.~Jiao, and S.~Torquato, \enquote{Computational
      design of anisotropic stealthy hyperuniform composites with engineered
      directional scattering properties,}  (2023).
    
    \bibitem{torquato_local_2003}
    S.~Torquato and F.~Stillinger, \enquote{Local density fluctuations,
      hyperuniformity, and order metrics,} {\protect\JournalTitle{Physical Review
      E}} \textbf{68}, 041113 (2003).
    
    \bibitem{torquato_hyperuniform_2018}
    S.~Torquato, \enquote{Hyperuniform states of matter,}
      {\protect\JournalTitle{Physics Reports}} \textbf{745}, 1--95 (2018).
    
    \bibitem{torquato_local_2022}
    S.~Torquato, M.~Skolnick, and J.~Kim, \enquote{Local order metrics for
      two-phase media across length scales*,} {\protect\JournalTitle{Journal of
      Physics A: Mathematical and Theoretical}} \textbf{55}, 274003 (2022).
    
    \bibitem{torquato_ensemble_2015}
    S.~Torquato, G.~Zhang, and F.~H. Stillinger, \enquote{Ensemble theory for
      stealthy hyperuniform disordered ground states,} {\protect\JournalTitle{Phys.
      Rev. X}} \textbf{5}, 021020 (2015).
    
    \bibitem{zhang_perfect_2016}
    G.~Zhang, F.~H. Stillinger, and S.~Torquato, \enquote{The perfect glass
      paradigm: Disordered hyperuniform glasses down to absolute zero,}
      {\protect\JournalTitle{Scientific Reports}} \textbf{6}, 36963 (2016).
    
    \bibitem{hexner_enhanced_2017}
    D.~Hexner, P.~Chaikin, and D.~Levine, \enquote{Enhanced hyperuniformity from
      random reorganization,} {\protect\JournalTitle{Proc. Natl. Acad. Sci.
      U.S.A.}} \textbf{114}, 4294--4299 (2017).
    
    \bibitem{oguz_hyperuniformity_2017}
    E.~O{\v g}uz, J.~Socolar, P.~Steinhardt, and S.~Torquato,
      \enquote{Hyperuniformity of quasicrystals,} {\protect\JournalTitle{Physical
      Review B}} \textbf{95}, 054119 (2017).
    
    \bibitem{ma_random_2017}
    Z.~Ma and S.~Torquato, \enquote{Random scalar fields and hyperuniformity,}
      {\protect\JournalTitle{Journal of Applied Physics}} \textbf{121}, 244904
      (2017).
    
    \bibitem{lopez_true_2018}
    C.~L{\'o}pez, \enquote{The true value of disorder,}
      {\protect\JournalTitle{Advanced Optical Materials}} \textbf{6}, 1800439
      (2018).
    
    \bibitem{yu_disordered_2018}
    S.~Yu, X.~Piao, and N.~Park, \enquote{Disordered potential landscapes for
      anomalous delocalization and superdiffusion of light,}
      {\protect\JournalTitle{ACS Photonics}} \textbf{5}, 1499--1505 (2018).
    
    \bibitem{wang_hyperuniformity_2018}
    J.~Wang, J.~M. Schwarz, and J.~D. Paulsen, \enquote{Hyperuniformity with no
      fine tuning in sheared sedimenting suspensions,}
      {\protect\JournalTitle{Nature Communications}} \textbf{9}, 2836 (2018).
    
    \bibitem{lei_hydrodynamics_2019}
    Q.-L. Lei and R.~Ni, \enquote{Hydrodynamics of random-organizing hyperuniform
      fluids,} {\protect\JournalTitle{Proceedings of the National Academy of
      Sciences of the United States of America}} \textbf{116}, 22983--22989 (2019).
    
    \bibitem{klatt_cloaking_2020}
    M.~A. Klatt, J.~Kim, and S.~Torquato, \enquote{Cloaking the underlying
      long-range order of randomly perturbed lattices,}
      {\protect\JournalTitle{Physical Review E}} \textbf{101}, 032118 (2020).
    
    \bibitem{nizam_dynamic_2021}
    {\"U}.~S. Nizam, G.~Makey, M.~Barbier, S.~S. Kahraman, E.~Demir, E.~E. Shafigh,
      S.~Galioglu, D.~Vahabli, S.~H{\"u}sn{\"u}gil, M.~H. G{\"u}ne{\c s},
      E.~Yelesti, and S.~Ilday, \enquote{Dynamic evolution of hyperuniformity in a
      driven dissipative colloidal system,} {\protect\JournalTitle{Journal of
      Physics: Condensed Matter}} \textbf{33}, 304002 (2021).
    
    \bibitem{torquato_multifunctional_2018}
    S.~Torquato and D.~Chen, \enquote{Multifunctional hyperuniform cellular
      networks: optimality, anisotropy and disorder,}
      {\protect\JournalTitle{Multifunct. Mater.}} \textbf{1}, 015001 (2018).
    
    \bibitem{kim_new_2019}
    J.~Kim and S.~Torquato, \enquote{New tessellation-based procedure to design
      perfectly hyperuniform disordered dispersions for materials discovery,}
      {\protect\JournalTitle{Acta Mater.}} \textbf{168}, 143--151 (2019).
    
    \bibitem{ghosh_generalized_2017}
    S.~Ghosh and J.~L. Lebowitz, \enquote{Generalized stealthy hyperuniform
      processes: maximal rigidity and the bounded holes conjecture,}
      {\protect\JournalTitle{Commun. Math. Phys.}} \textbf{363}, 97--110 (2018).
    
    \bibitem{brauchart_hyperuniform_2019}
    J.~S. Brauchart, P.~J. Grabner, and W.~Kusner, \enquote{Hyperuniform point sets
      on the sphere: Deterministic aspects,} {\protect\JournalTitle{Constructive
      Approximation}} \textbf{50}, 45--61 (2019).
    
    \bibitem{torquato_hidden_2019}
    S.~Torquato, G.~Zhang, and M.~D. {Courcy-Ireland}, \enquote{Hidden multiscale
      order in the primes,} {\protect\JournalTitle{Journal of Physics A:
      Mathematical and Theoretical}} \textbf{52}, 135002 (2019).
    
    \bibitem{lacroix_intermedidate_2019}
    B.~Lacroix-A-Chez-Toine, J.~A.~M. Garz{\'o}n, C.~S.~H. Calva, I.~P. Castillo,
      A.~Kundu, S.~N. Majumdar, and G.~Schehr, \enquote{Intermediate deviation
      regime for the full eigenvalue statistics in the complex ginibre ensemble,}
      {\protect\JournalTitle{Physical Review E}} \textbf{100}, 012137 (2019).
    
    \bibitem{jiao_avian_2014}
    Y.~Jiao, T.~Lau, H.~Hatzikirou, M.~{Meyer-Hermann}, C.~Corbo, and S.~Torquato,
      \enquote{Avian photoreceptor patterns represent a disordered hyperuniform
      solution to a multiscale packing problem,} {\protect\JournalTitle{Physical
      Review E}} \textbf{89}, 022721 (2014).
    
    \bibitem{mayer_how_2015}
    A.~Mayer, V.~Balasubramanian, T.~Mora, and A.~M. Walczak, \enquote{How a
      well-adapted immune system is organized,} {\protect\JournalTitle{Proceedings
      of the National Academy of Sciences of the United States of America}}
      \textbf{112}, 5950--5955 (2015).
    
    \bibitem{klatt_characterization_2018}
    M.~A. Klatt and S.~Torquato, \enquote{Characterization of maximally random
      jammed sphere packings. \{III. T\}ransport and electromagnetic properties via
      correlation functions,} {\protect\JournalTitle{Physical Review E}}
      \textbf{97}, 012118 (2018).
    
    \bibitem{froufe-perez_bandgap_2023}
    L.~S. {Froufe-P{\'e}rez}, G.~J. Aubry, F.~Scheffold, and S.~Magkiriadou,
      \enquote{Bandgap fluctuations and robustness in two-dimensional hyperuniform
      dielectric materials,} {\protect\JournalTitle{Opt. Express, OE}} \textbf{31},
      18509--18515 (2023).
    
    \bibitem{ong_control_2023}
    Z.-Y. Ong, \enquote{Control of wave scattering for robust coherent transmission
      in a disordered medium,}  (2023).
    
    \bibitem{zachary_hyperuniformity_2009}
    C.~Zachary and S.~Torquato, \enquote{Hyperuniformity in point patterns and
      two-phase random heterogeneous media,} {\protect\JournalTitle{Journal of
      Statistical Mechanics: Theory and Experiment}} \textbf{2009}, P12015 (2009).
    
    \bibitem{uche_constraints_2004}
    O.~Uche, F.~Stillinger, and S.~Torquato, \enquote{Constraints on collective
      density variables: Two dimensions,} {\protect\JournalTitle{Physical Review
      E}} \textbf{70}, 046122 (2004).
    
    \bibitem{batten_classical_2008}
    R.~Batten, F.~Stillinger, and S.~Torquato, \enquote{Classical disordered ground
      states: Super-ideal gases and stealth and equi-luminous materials,}
      {\protect\JournalTitle{J. Appl. Phys.}} \textbf{104}, 033504 (2008).
    
    \bibitem{zhang_ground_2015}
    G.~Zhang, F.~Stillinger, and S.~Torquato, \enquote{Ground states of stealthy
      hyperuniform potentials: I. entropically favored configurations,}
      {\protect\JournalTitle{Phys. Rev. E}} \textbf{92}, 022119 (2015).
    
    \bibitem{zhang_transport_2016}
    G.~Zhang, F.~Stillinger, and S.~Torquato, \enquote{Transport, geometrical, and
      topological properties of stealthy disordered hyperuniform two-phase
      systems,} {\protect\JournalTitle{J. Chem. Phys.}} \textbf{145}, 244109
      (2016).
    
    \bibitem{zhou_ultrabroadband_2020}
    W.~Zhou, Y.~Tong, X.~Sun, and H.~K. Tsang, \enquote{Ultra-broadband
      hyperuniform disordered silicon photonic polarizers,}
      {\protect\JournalTitle{IEEE J. Sel. Top. Quantum Electron.}} \textbf{26},
      1--9 (2020).
    
    \bibitem{sheremet_absorption_2020}
    A.~Sheremet, R.~Pierrat, and R.~Carminati, \enquote{Absorption of scalar waves
      in correlated disordered media and its maximization using stealth
      hyperuniformity,} {\protect\JournalTitle{Phys. Rev. A}} \textbf{101}, 053829
      (2020).
    
    \bibitem{stogryn_electromagnetic_1974}
    A.~Stogryn, \enquote{Electromagnetic scattering by random dielectric constant
      fluctuations in a bounded medium,} {\protect\JournalTitle{Radio Science}}
      \textbf{9}, 509--518 (1974).
    
    \bibitem{tsang_theory_1977}
    L.~Tsang and J.~A. Kong, \enquote{Theory for thermal microwave emission from a
      bounded medium containing spherical scatterers,} {\protect\JournalTitle{J.
      Appl. Phys.}} \textbf{48}, 3593--3599 (1977).
    
    \bibitem{tsang_scattering_1981}
    L.~Tsang and J.~A. Kong, \enquote{Scattering of electromagnetic waves from
      random media with strong permittivity fluctuations,}
      {\protect\JournalTitle{Radio Sci.}} \textbf{16}, 303--320 (1981).
    
    \bibitem{sihvola_electromagnetic_1999}
    A.~Sihvola, \emph{Electromagnetic Mixing Formulas and Applications} (IET
      Digital Library, London, 1999).
    
    \bibitem{tatarskii_effects_1971}
    V.~I. Tatarskii, \emph{The Effects of the Turbulent Atmosphere on Wave
      Propagation} (Jerusalem: Israel Program for Scientific Translations,
      Springfield, 1971).
    
    \bibitem{silveirinha_design_2007}
    M.~Silveirinha and N.~Engheta, \enquote{Design of matched zero-index
      metamaterials using nonmagnetic inclusions in epsilon-near-zero media,}
      {\protect\JournalTitle{Physical Review B}} \textbf{75}, 075119 (2007).
    
    \bibitem{rytov_electromagnetic_1956}
    S.~Rytov, \enquote{Electromagnetic properties of a finely stratified medium,}
      {\protect\JournalTitle{Sov. Phys. JEPT}} \textbf{2}, 466--475 (1956).
    
    \bibitem{sjoberg_exact_2006}
    D.~Sj{\"o}berg, \enquote{Exact and asymptotic dispersion relations for
      homogenization of stratified media with two phases,}
      {\protect\JournalTitle{J. Electromagn. Waves Appl.}} \textbf{20}, 781--792
      (2006).
    
    \bibitem{maurel_effective_2008}
    A.~Maurel and V.~Pagneux, \enquote{Effective propagation in a perturbed
      periodic structure,} {\protect\JournalTitle{Phys. Rev. B}} \textbf{78},
      052301 (2008).
    
    \bibitem{chebykin_nonlocal_2011}
    A.~V. Chebykin, A.~A. Orlov, A.~V. Vozianova, S.~I. Maslovski, {\relax Yu}.~S.
      Kivshar, and P.~A. Belov, \enquote{Nonlocal effective medium model for
      multilayered metal-dielectric metamaterials,} {\protect\JournalTitle{Phys.
      Rev. B}} \textbf{84}, 115438 (2011).
    
    \bibitem{popov_operator_2016}
    V.~Popov, A.~V. Lavrinenko, and A.~Novitsky, \enquote{Operator approach to
      effective medium theory to overcome a breakdown of maxwell garnett
      approximation,} {\protect\JournalTitle{Phys. Rev. B}} \textbf{94}, 085428
      (2016).
    
    \bibitem{merzlikin_homogenization_2020}
    A.~M. Merzlikin and R.~S. Puzko, \enquote{Homogenization of maxwell's equations
      in a layered system beyond the static approximation,}
      {\protect\JournalTitle{Sci Rep}} \textbf{10}, 15783 (2020).
    
    \bibitem{wen_nonlocal_2021}
    Z.~Wen, H.~Xu, W.~Zhao, Z.~Zhou, X.~Li, S.~Li, J.~Zhou, Y.~Sun, N.~Dai, and
      J.~Hao, \enquote{Nonlocal effective-medium theory for periodic multilayered
      metamaterials,} {\protect\JournalTitle{J. Opt.}} \textbf{23}, 065103 (2021).
    
    \bibitem{siqueira_method_1996}
    P.~Siqueira and K.~Sarabandi, \enquote{Method of moments evaluation of the
      two-dimensional quasi-crystalline approximation,} {\protect\JournalTitle{IEEE
      Trans. Antennas Propag.}} \textbf{44}, 1067--1077 (1996).
    
    \bibitem{odeh_optical_2021}
    M.~Odeh, M.~Dupr{\'e}, K.~Kim, and B.~Kant{\'e}, \enquote{Optical response of
      jammed rectangular nanostructures,} {\protect\JournalTitle{Nanophotonics}}
      \textbf{10}, 705--711 (2021).
    
    \bibitem{keller_stochastic_1964}
    J.~B. Keller, \enquote{Stochastic equations and wave propagation in random
      media,} {\protect\JournalTitle{Proceedings of Symposia in Applied
      Mathematics}} \textbf{16}, 145--170 (1964).
    
    \bibitem{ruppin_evaluation_2000}
    R.~Ruppin, \enquote{Evaluation of extended maxwell-garnett theories,}
      {\protect\JournalTitle{Optics Communications}} \textbf{182}, 273--279 (2000).
    
    \bibitem{rechtsman_effective_2008}
    M.~C. Rechtsman and S.~Torquato, \enquote{Effective dielectric tensor for
      electromagnetic wave propagation in random media,} {\protect\JournalTitle{J.
      Appl. Phys.}} \textbf{103}, 084901 (2008).
    
    \bibitem{torquato_nonlocal_2021}
    S.~Torquato and J.~Kim, \enquote{Nonlocal effective electromagnetic wave
      characteristics of composite media: Beyond the quasistatic regime,}
      {\protect\JournalTitle{Phys. Rev. X}} \textbf{11}, 021002 (2021).
    
    \bibitem{frisch_probabilistic_1968}
    U.~Frisch, \emph{Probabilistic Methods in Applied Mathematics}, vol. I and II
      (Academic Press, New York, 1968).
    
    \bibitem{caze_diagrammatic_2015}
    A.~Caz{\'e} and J.~C. Schotland, \enquote{Diagrammatic and asymptotic
      approaches to the origins of radiative transport theory: tutorial,}
      {\protect\JournalTitle{J. Opt. Soc. Am. A}} \textbf{32}, 1475 (2015).
    
    \bibitem{kim_effective_2023}
    J.~Kim and S.~Torquato, \enquote{Effective electromagnetic wave properties of
      disordered stealthy hyperuniform layered media beyond the quasistatic
      regime,} {\protect\JournalTitle{Optica}} \textbf{10}, 965 (2023).
    
    \bibitem{jackson_classical_1999}
    J.~D. Jackson, \emph{Classical Electrodynamics} (John Wiley \& Sons, Inc., New
      York, 1999), 3rd ed.
    
    \bibitem{torquato_inverse_2009}
    S.~Torquato, \enquote{Inverse optimization techniques for targeted
      self-assembly,} {\protect\JournalTitle{Soft Matter}} \textbf{5}, 1157--1173
      (2009).
    
    \bibitem{torquato_random_2002}
    S.~Torquato, \emph{Random Heterogeneous Materials: Microstructure and
      Macroscopic Properties}, no.~16 in {\emph{Interdisciplinary Applied
      Mathematics}} (Springer Science \& Business Media, New York, 2002).
    
    \bibitem{torquato_microstructure_1982}
    S.~Torquato and G.~Stell, \enquote{Microstructure of two-phase random media.
      \{I\}. \{T\}he \$n\$-point probability functions,} {\protect\JournalTitle{J.
      Chem. Phys.}} \textbf{77}, 2071--2077 (1982).
    
    \bibitem{debye_scattering_1957}
    P.~Debye, H.~R. Anderson~Jr, and H.~Brumberger, \enquote{Scattering by an
      inhomogeneous solid. ii. the correlation function and its application,}
      {\protect\JournalTitle{J. Appl. Phys.}} \textbf{28}, 679--683 (1957).
    
    \bibitem{torquato_perspective_2018}
    S.~Torquato, \enquote{Perspective: Basic understanding of condensed phases of
      matter via packing models,} {\protect\JournalTitle{J. Chem. Phys.}}
      \textbf{149}, 020901 (2018).
    
    \bibitem{torquato_hyperuniformity_2016}
    S.~Torquato, \enquote{Hyperuniformity and its generalizations,}
      {\protect\JournalTitle{Phys. Rev. E}} \textbf{94}, 022122 (2016).
    
    \bibitem{torquato_structural_2021}
    S.~Torquato, \enquote{Structural characterization of many-particle systems on
      approach to hyperuniform states,} {\protect\JournalTitle{Physical Review E}}
      \textbf{103}, 052126 (2021).
    
    \bibitem{stanley_introduction_1987}
    H.~E. Stanley, \emph{Introduction to phase transitions and critical phenomena},
      International series of monographs on physics (Oxford university press, New
      York Oxford, 1987).
    
    \bibitem{binney_theory_2002}
    J.~J. Binney and J.~J. Binney, eds., \emph{The theory of critical phenomena: an
      introduction to the renormalization group}, Oxford science publications
      (Oxford Univ. Press, Oxford, 2002), reprinted with corr ed.
    
    \bibitem{mandelbrot_fractal_1982}
    B.~B. Mandelbrot, \emph{The fractal geometry of nature} (W.H. Freeman, San
      Francisco, 1982).
    
    \bibitem{oguz_hyperuniformity_2019}
    E.~C. O{\u g}uz, J.~E.~S. Socolar, P.~J. Steinhardt, and S.~Torquato,
      \enquote{Hyperuniformity and anti-hyperuniformity in one-dimensional
      substitution tilings,} {\protect\JournalTitle{Acta Crystallographica Section
      A Foundations and Advances}} \textbf{75}, 3--13 (2019).
    
    \bibitem{mackay_strongpropertyfluctuation_2000}
    T.~G. Mackay, A.~Lakhtakia, and W.~S. Weiglhofer,
      \enquote{Strong-property-fluctuation theory for homogenization of
      bianisotropic composites: Formulation,} {\protect\JournalTitle{Phys. Rev. E}}
      \textbf{62}, 6052--6064 (2000).
    
    \bibitem{bruggeman_berechnung_1935}
    D.~A.~G. Bruggeman, \enquote{Berechnung verschiedener physikalischer konstanten
      von heterogenen substanzen. i. dielektrizit\"atskonstanten und
      leitf\"ahigkeiten der mischk\"orper aus isotropen substanzen,}
      {\protect\JournalTitle{Annalen der Physik}} \textbf{416}, 636--664 (1935).
    
    \bibitem{sheppard_greenfunction_2014}
    C.~J.~R. Sheppard, S.~S. Kou, and J.~Lin, \enquote{The green-function transform
      and wave propagation,} {\protect\JournalTitle{Frontiers in Physics}}
      \textbf{2} (2014).
    
    \bibitem{kim_extraordinary_2023}
    J.~Kim and S.~Torquato, \enquote{Extraordinary optical and transport properties
      of disordered stealthy hyperuniform two-phase media,}  (2023). In
      preparation.
    
    \bibitem{anderson_absence_1958}
    P.~W. Anderson, \enquote{Absence of diffusion in certain random lattices,}
      {\protect\JournalTitle{Phys. Rev.}} \textbf{109}, 1492--1505 (03/01/ 1958).
    
    \bibitem{mcgurn_anderson_1993}
    A.~R. McGurn, K.~T. Christensen, F.~M. Mueller, and A.~A. Maradudin,
      \enquote{Anderson localization in one-dimensional randomly disordered optical
      systems that are periodic on average,} {\protect\JournalTitle{Phys. Rev. B}}
      \textbf{47}, 13120--13125 (1993).
    
    \bibitem{sheng_introduction_2006}
    P.~Sheng, \emph{Introduction to wave scattering, localization, and mesoscopic
      phenomena}, no.~88 in Springer series in materials science (Springer, Berlin
      ; New York, 2006), 2nd ed.
    
    \bibitem{aegerter_coherent_2009}
    C.~M. Aegerter and G.~Maret, \enquote{Coherent backscattering and anderson
      localization of light,} in \emph{Progress in Optics,}  vol.~52 (Elsevier,
      2009), pp. 1--62.
    
    \bibitem{wiersma_disordered_2013}
    D.~S. Wiersma, \enquote{Disordered photonics,} {\protect\JournalTitle{Nature
      Photonics}} \textbf{7}, 188--196 (2013).
    
    \bibitem{kim_methodology_2019}
    J.~Kim and S.~Torquato, \enquote{Methodology to construct large realizations of
      perfectly hyperuniform disordered packings,} {\protect\JournalTitle{Phys.
      Rev. E}} \textbf{99}, 052141 (2019).
    
    \bibitem{zhang_can_2017}
    G.~Zhang, F.~H. Stillinger, and S.~Torquato, \enquote{Can exotic disordered
      ``stealthy" particle configurations tolerate arbitrarily large holes?}
      {\protect\JournalTitle{Soft Matter}} \textbf{13}, 6197--6207 (2017).
    
    \bibitem{yaghjian_electric_1980}
    A.~Yaghjian, \enquote{Electric dyadic green's functions in the source region,}
      {\protect\JournalTitle{Proc. IEEE}} \textbf{68}, 248--263 (1980).
    
    \bibitem{Kl_local_2023}
    M.~A. Klatt, P.~J. Steinhardt, and S.~Torquato, \enquote{~~,}
      {\protect\JournalTitle{preprint}}  (2023). In preparation.
    
    \bibitem{affinito_new_1996}
    J.~D. Affinito, M.~E. Gross, C.~A. Coronado, G.~L. Graff, I.~N. Greenwell, and
      P.~M. Martin, \enquote{A new method for fabricating transparent barrier
      layers,} {\protect\JournalTitle{Thin Solid Films}} \textbf{290--291}, 63--67
      (1996).
    
    \bibitem{lee_LayerbyLayer_2001}
    S.-S. Lee, J.-D. Hong, C.~H. Kim, K.~Kim, J.~P. Koo, and K.-B. Lee,
      \enquote{Layer-by-layer deposited multilayer assemblies of ionene-type
      polyelectrolytes based on the spin-coating method,}
      {\protect\JournalTitle{Macromolecules}} \textbf{34}, 5358--5360 (2001).
    
    \bibitem{zhao_assembly_2018}
    K.~Zhao and T.~G. Mason, \enquote{Assembly of colloidal particles in solution,}
      {\protect\JournalTitle{Reports on Progress in Physics}} \textbf{80}, 126601
      (2018).
    
    \bibitem{tumbleston_continuous_2015}
    J.~R. Tumbleston, D.~Shirvanyants, N.~Ermoshkin, R.~Janusziewicz, A.~R.
      Johnson, D.~Kelly, K.~Chen, R.~Pinschmidt, J.~P. Rolland, A.~Ermoshkin, E.~T.
      Samulski, and J.~M. DeSimone, \enquote{Continuous liquid interface production
      of 3d objects,} {\protect\JournalTitle{Science}} \textbf{347}, 1349--1352
      (2015).
    
    \bibitem{taflove_advances_2013}
    A.~Taflove, S.~G. Johnson, and A.~Oskooi, \emph{Advances in FDTD Computational
      Electrodynamics: Photonics and Nanotechnology} (Artech House, Boston, 2013).
    
    \bibitem{oskooi_meep_2010}
    A.~F. Oskooi, D.~Roundy, M.~Ibanescu, P.~Bermel, J.~D. Joannopoulos, and S.~G.
      Johnson, \enquote{Meep: A flexible free-software package for electromagnetic
      simulations by the fdtd method,} {\protect\JournalTitle{Comput. Phys.
      Commun.}} \textbf{181}, 687--702 (2010).
    
    \end{thebibliography}


\end{document}